\documentclass{article}

\usepackage[preprint]{neurips_2026}

\usepackage[utf8]{inputenc} 
\usepackage[T1]{fontenc}    
\usepackage{graphicx}
\usepackage{subcaption}
\usepackage[inkscapelatex=false]{svg}
\usepackage{ulem}
\usepackage{amsmath}
\usepackage{mathtools}
\usepackage{multirow}
\usepackage{hyperref}       
\usepackage{wrapfig}
\usepackage{makecell}
\setcitestyle{round}
\hypersetup{
   colorlinks=true,
   linkcolor=[RGB]{30, 30, 180},
   citecolor=[RGB]{30, 30, 180},
   urlcolor=magenta,
   pdfborder=0 0 0,
   pdftitle={},
   pdfsubject={},
   pdfkeywords={},
   pdfauthor={},%
   pdfstartview=FitH
}
\usepackage[capitalize,noabbrev]{cleveref}
\usepackage[nolist,nohyperlinks]{acronym}
\usepackage{pifont}
\usepackage{booktabs}
\usepackage[table]{xcolor} 
\usepackage[inline]{enumitem}
\usepackage{xspace}
\usepackage[inline]{enumitem}

\usepackage{array}
\usepackage{amsmath}
\usepackage{amssymb}
\usepackage{mathtools}
\usepackage{amsthm}
\usepackage{bm}
\usepackage{xspace}

\newcommand{\ourmethod}{GyroFlow}
\newcommand{\ourfid}{FGyD}

\newcommand\Bb{\bm{b}}
\newcommand\Bc{\bm{c}}

\newcommand\Bk{\bm{k}}

\newcommand\Bv{\bm{v}}

\newcommand\BB{\bm{B}}

\newcommand\BE{\bm{E}}

 \newcommand{\cD}{\mathcal{D}}

\makeatletter
\newcommand{\dlmf}[1]{%
\citep[%
  \def\nextitem{\def\nextitem{, }}%
  \@for \el:=#1\do{\nextitem\href{http://dlmf.nist.gov/\el}{(\el)}}%
]{olver_nist_2010}%
}
\makeatother

\definecolor{paramRLT}{HTML}{CC7B7B}    
\definecolor{paramRLn}{HTML}{CCA265}    
\definecolor{paramQ}{HTML}{65AA85}      
\definecolor{paramShat}{HTML}{8878B8}   
\definecolor{termKinetic}{HTML}{B088A8}  
\definecolor{termDrive}{HTML}{58A8A0}    
\definecolor{termNL}{HTML}{CC9072}       
\definecolor{termDiss}{HTML}{6890B5}     

\newcommand{\paramph}{\vphantom{R/L_T\hat{s}}}
\newcommand{\paramRLTbox}{\colorbox{paramRLT!20}{$\paramph R/L_T$}}
\newcommand{\paramRLnbox}{\colorbox{paramRLn!20}{$\paramph R/L_n$}}
\newcommand{\paramQbox}{\colorbox{paramQ!20}{$\paramph q$}}
\newcommand{\paramShatbox}{\colorbox{paramShat!20}{$\paramph \hat{s}$}}

\newcommand{\textbfp}[1]{\vspace{0.5em}\noindent\textbf{#1.}}

\newcommand{\cmark}{\cellcolor{green!20}\ding{51}} 
\newcommand{\xmark}{\cellcolor{red!20}\ding{55}}   

\title{
  A Shortcut to Statistically Steady-State Turbulence with Flow Matching
}

\author{
  \vspace{-2em}\\
  \textbf{Gianluca Galletti} \thanks{Equal contribution, $\:$ corresponding author: \texttt{galletti@ml.jku.at}} $\:^{1}$ \hspace{16px}%
  \textbf{Gerald Gutenbrunner} \footnotemark[1] $\:^{1}$ \\[0.3em]
  \textbf{William Hornsby} $^{2}$ \hspace{16px}%
  \textbf{Lorenzo Zanisi} $^{2}$ \hspace{16px}%
  \textbf{Naomi Carey} $^{2}$ \\[0.3em]
  \textbf{Stanislas Pamela} $^{2}$ \hspace{16px}%
  \textbf{Johannes Brandstetter} $^{1,3}$ \hspace{16px}%
  \textbf{Fabian Paischer} $^{1,3}$ \vspace{0.5em}\\[0.3em]
  {$^1$~Institute for Machine Learning, JKU Linz}\\
  {$^2$~United Kingdom Atomic Energy Authority, Culham campus}\\
  {$^3$~Mistral AI}
  \\[0.5em]
  \href{https://github.com/ml-jku/neural-gyrokinetics}{\raisebox{-0.3em}{\includegraphics[width=1.15em]{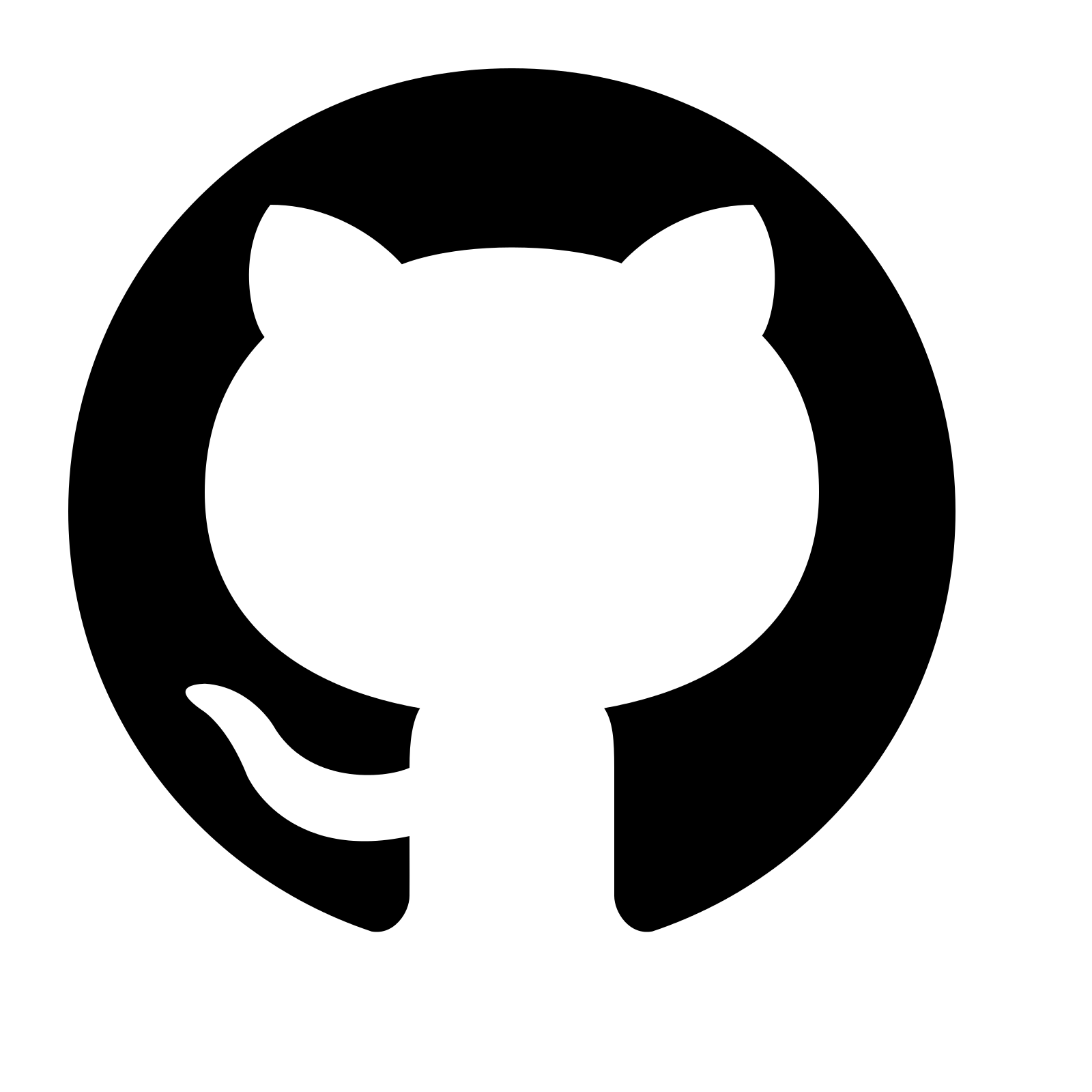}} \texttt{ml-jku/neural-gyrokinetics}}\hspace{1em}
\href{https://huggingface.co/datasets/gerkone/cbc-gyroswin-256traj}{\raisebox{-0.3em}{\includegraphics[width=1.15em]{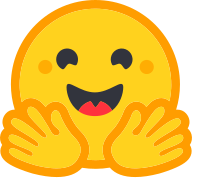}} \texttt{gerkone/cbc-gyroswin-256traj}}
\vspace{-12px}
}

\begin{document}

\maketitle

\begin{abstract}
    Many nonlinear physical systems exhibit an initial transient phase in which perturbations grow before nonlinear interactions lead to a statistically steady state.
    While this saturated regime is of primary interest, direct numerical simulations must resolve the full transient dynamics before reaching it, incurring significant computational cost.
    In Computational Fluid Dynamics, reduced-order approaches such as Large Eddy Simulation mitigate computational cost by modeling small-scale dynamics, enabling tractable approximations of turbulent flows.
    In contrast, for systems such as gyrokinetics, comparably effective closures for the full dynamics are not generally available, and high-fidelity simulations remain necessary.
    Existing surrogate modeling approaches for these systems are autoregressive, hence they suffer from accumulating error.
    We instead propose to bypass explicit time evolution by directly modeling the distribution of saturated states under an ergodicity assumption, stating that ensemble averages over samples are equivalent to time averages of a single long simulation. 
    We introduce \ourmethod, a latent generative model that directly estimates steady-state statistics of gyrokinetic turbulence in 5D phase space, without resolving the transient phase.
    \ourmethod{} generates saturated snapshots from noise, conditioned on dimensionless operating parameters and outperforms autoregressive, reduced-order, and other generative approaches, while providing substantial speedup.
    To evaluate generation quality we propose \ourfid{}, a distributional metric computed in the latent space of a pretrained gyrokinetic model, and show that it correlates with downstream flux accuracy and solver convergence.
    Finally, \ourmethod{} can be used to warm-start the numerical code used to produce the data.
\end{abstract}


\section{Introduction}
\label{sec:intro}

Numerical simulations of turbulent multiscale systems are a cornerstone of modern science and engineering, from aerospace and automotive design \citep{slotnick2014cfd2030} to interstellar medium modeling \citep{Federrath2012} and energy systems, where they predict the turbulent transport that determines confinement in nuclear fusion power plant concepts such as STEP \citep{Kennedy2023} and underpin the design of combustion engines and gas turbines \citep{pitsch2006large}.
After an initial transient, these systems relax to a statistically steady regime from which quantities of interest for design or scientific analysis, such as heat and momentum transport in plasmas, are typically obtained via time averaging.
The initial transient phase can be long and costly to compute, yet it is strictly necessary to simulate to reach the statistically steady state regime and therefore the quantities of interest. 

In Computational Fluid Dynamics, substantial speedups can be obtained via Large Eddy Simulation, where dissipation at small scales is modeled instead of resolved explicitly, with many analytical closures available \citep{pope2000turbulent}.
However, such reductions collapse when no tractable closure is available or the observables of interest are themselves statistics of multiscale fluctuations.
Gyrokinetics is a prominent example of this problem class, and one of the most challenging simulation problems in engineering.
It models the turbulent transport that governs confinement in magnetic fusion plasmas, and therefore the viability of fusion as an energy source.
Kinetic effects in the plasma core 
and the low collisionality between particles require a 5D phase-space description, rather than a fluid approximation. 
An approach similar to LES in gyrokinetics is notoriously difficult, because of the bi-directional energy cascade where smaller scales drive an inverse transfer that generates large-scale structures such as zonal flows, which in turn suppress the small scales.
Therefore reduced models only exist via data-driven semi-empirical saturation rules \citep{Staebler2007,bourdelle_qualikiz_2007}, which are brittle in strongly turbulent and power plant-relevant regimes \citep{dimits_shift_2000,Bourdelle_validity_2008,Kiefer_quasilinear_2021}.
Reliable predictions for fusion power plants require multiple expensive, time-dependent nonlinear simulations, including a transient ramp-up phase during which unstable modes grow before nonlinear couplings and zonal flows saturate the turbulence.
%

Machine learning based surrogate models of gyrokinetics offer a fruitful alternative.
Existing approaches sit on two opposite sides of a trade-off.
Autoregressive neural surrogates \citep{paischer2025gyroswin} retain the 5D dynamics but integrate through the transient, accumulating rollout error as they proceed.
Quasilinear reduced-order models \citep{Bourdelle2015,Citrin2017,Staebler2007,Staebler2010} and classical surrogates \citep{Hornsby2024,plassche_qlknn_2020} bypass the transient entirely, but discard the nonlinear physics.

In this work we attempt a different approach that brings together the best of both worlds.
We leverage the fact that gyrokinetic turbulence can be treated as approximately ergodic, as
time-averaged observables converge to their ensemble averages over the underlying steady-state distribution after a sufficiently long time.
These statistics are primarily determined by the dimensionless operating parameters. 
Therefore, we propose \ourmethod{}, a latent flow matching \citep{lipmanflow,dao2023flowmatchinglatentspace} model trained to sample turbulent 5D snapshots conditioned on operating parameters.
This enables recovering the ensemble average in a single forward pass without the need for autoregressive rollouts or reducing assumptions.
To evaluate generation of 5D plasma data we propose \ourfid, a Fr\'echet distance in the latent space of a pretrained gyrokinetic surrogate \citep{paischer2025gyroswin}, whose features encode mode structure, amplitude, and zonal-flow content and verify that it correlates with downstream flux accuracy and the convergence time of warm-started solvers.

Our contributions are summarized as follows.
\begin{enumerate}[label={\ding{\numexpr181+\value{enumi}\relax}}, labelsep=0.4em, leftmargin=3.0em, rightmargin=3.0em, itemsep=0.05em, topsep=0.0em]
    \item We introduce \ourmethod{}, a generative model of saturated 5D gyrokinetic turbulence which bypasses the costly transient, conditioned on operating parameters.
    \item We introduce \ourfid{} as a distributional metric defined in the latent space of GyroSwin \citep{paischer2025gyroswin}, and show that it correlates with downstream performance.
    \item We demonstrate that generated fields can be effectively used to warm-start 
    numerical solvers, reducing their time-to-convergence to the saturated regime.
\end{enumerate}

\begin{figure}[t]
\centering
\includegraphics[width=\textwidth]{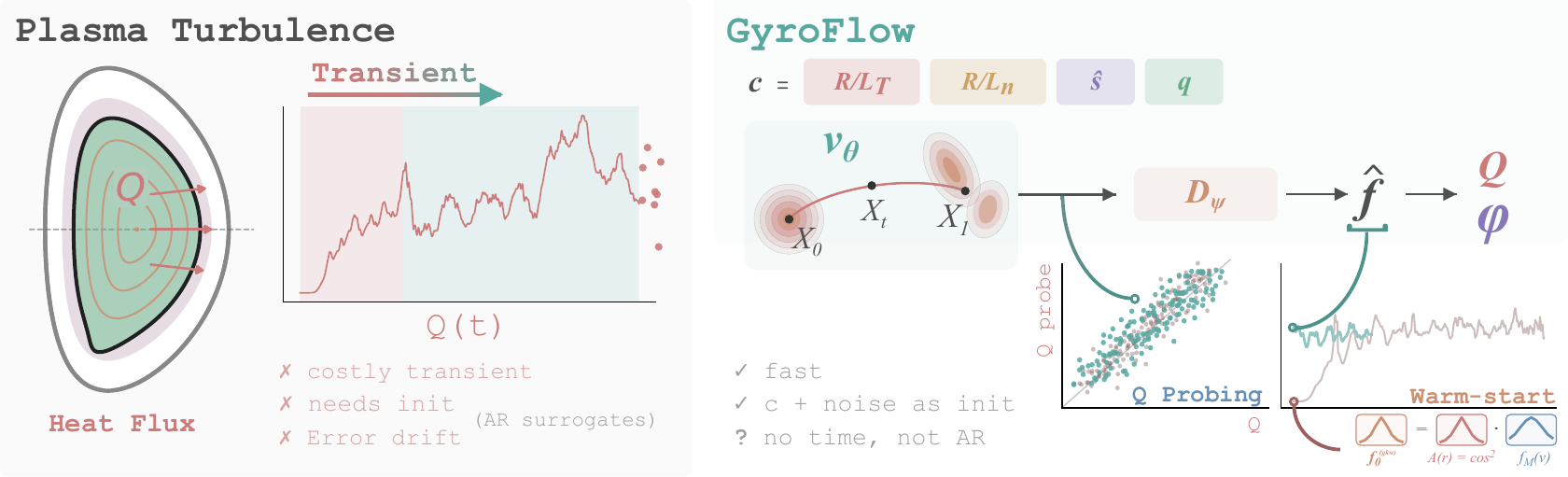}
\caption{
\textbf{Left:} In a Tokamak the heat flux $Q(t)$ is the radial transport of thermal energy across the nested magnetic flux surfaces, driven by turbulence. 
The transient of a gyrokinetic simulation traverses an initial ramp-up phase before converging to a statistically steady state.
\textbf{Right:} \ourmethod{} assumes ergodicity and uses flow matching to sample directly from the saturated distribution, instead of traversing the transient.
Downstream quantities such as average heat flux can be obtained either by latent probing or reconstruction of the 5D phase space.
}
\label{fig:figure1}
\end{figure}

\section{The (Un)avoidable Cost of Gyrokinetics}
\label{sec:background}




Gyrokinetic simulation codes, such as GKW \citep{PEETERS_GKW_2009}, evolve a 5D distribution function $f(k_x, k_y, s, v_\parallel, \mu;\, t)$ of charged particles in a magnetised plasma.
Turbulence in a plasma is driven by microinstabilities that draw free energy from the radial gradients of temperature and density, giving rise to, e.g., ion-temperature-gradient \citep[ITG]{coppi_1967_itg} modes.
These instabilities are responsible for the transport of heat and particles, which is the central quantity governing the performance of magnetically confined fusion devices.

A gyrokinetic simulation proceeds through two distinct phases. In the linear phase, perturbations to the distribution function grow exponentially as the microinstabilities amplify.
Once the fluctuation amplitudes become sufficiently large, nonlinear mode coupling sets in and energy is redistributed across scales which causes emergence of zonal flows \citep{Itoh_ZF_06}.
Zonal flows act as a self-regulation mechanism, because they shear apart the turbulent eddies at small scales, effectively suppressing transport and saturating the instability growth.
The system transitions into a saturated regime characterised by a dynamic equilibrium between instability drive and zonal flow regulation.

Simulations with identical parameters but different initial conditions converge to the same steady-state statistics. 
This indicates that gyrokinetic turbulence is approximately ergodic, meaning that time averages along a single realisation equal ensemble averages, defining a unique statistical steady state.
Together, these properties imply that transport-relevant statistics, such as mean and fluctuating heat and particle fluxes, are fully characterised by the operating parameters alone.
No information about the final statistical state is carried by the ramp-up trajectory, therefore it can be considered pure computational overhead.

In practice, the linear buildup phase is so long and costly that the ensemble average is impractical, and therefore time averages along a single simulation are taken. 
However, the cost incurred by traversing the transient of a single gyrokinetic simulation is only the tip of the iceberg.
Integrated modeling of plasma in a tokamak \citep{VanMulders2021,citrin2024toraxfastdifferentiabletokamak,Bourdelle2025} compounds this cost by several orders of magnitude, since a turbulent-transport prediction is required at every radial point and every time slice of the discharge, requiring many thousands of evaluations for fusion power plant design \citep{zanisi2025}.
While nonlinear gyrokinetics provide the high-fidelity information needed to de-risk these designs, their computational cost makes such integrated predictions currently infeasible.

Based on the observation that the saturated distribution depends only on the operating parameters and is accessible through independent samples under the ergodicity assumption, we aim to replace temporal simulations with recent generative modeling approaches.
That is, a model that maps operating parameters directly to realisations of the saturated state, bypassing both the ramp-up and the need for long time-averaging trajectories.

\section{Related Works}
\label{sec:related_work}
\textbfp{Surrogates for gyrokinetics}
There have been multiple attempts to develop surrogates for core turbulent transport as modelled by gyrokinetics.
Quasilinear reduced-order models such as QuaLiKiz \citep{Bourdelle2015,Citrin2017} and TGLF \citep{Staebler2007,Staebler2010} replace the nonlinear term in \cref{eq:gyrovlasov} with a saturation rule fitted to nonlinear simulations and combine independent $k_y$-mode contributions via a weighting function \citep{Staebler_quasilinear_2024}.
They evolve reduced linear simulations that resolve each mode independently, hence reducing the dimensionality to 3D, but neglect nonlinear phenomena.
Therefore, they are generally less accurate near stability boundaries and in strongly driven regimes \citep{dimits_shift_2000,Bourdelle_validity_2008}.
Classical surrogates, namely Gaussian process regression \citep{Hornsby2024} and multilayer perceptrons \citep{plassche_qlknn_2020,Citrin_qlknn_2023,Zanisi2024}, map operating parameters directly to scalar fluxes.
They are efficient, but inherit the same limitations as reduced-order models.

Recent work targets full gyrokinetic surrogates.
\citet{narita_toward_2022} apply CNNs to 2D wavenumber slices for flux and time-to-saturation prediction, \citet{mitsuru_multimodal_2023} extend this with multimodal inputs, and \citet{Wan_itgtransfer_2025} study cross-fidelity transfer in a reduced 1D space.
GyroSwin \citep{paischer2025gyroswin} is the first surrogate operating on the full 5D phase space with orders-of-magnitude per-step speedup and state of the art accuracy.
It still depends on an initially evolved saturated state from a numerical code and also traverses the transient step by step.
Therefore it suffers from error accumulation affecting the time averages of the quantities of interest.

\textbfp{Generative models for scientific simulation}
Denoising diffusion models \citep{ho2020denoising,song2020improved,dhariwal2021diffusion} sample from a target distribution by inverting a forward noising process and dominate content generation across images, video, and audio \citep{yang2025diffusionmodelscomprehensivesurvey}. Deterministic samplers \citep{song2021denoising,karras2022_Elucidating} and continuous-time ODE reformulations (flow matching \citep{lipmanflow}, stochastic interpolants \citep{albergo2023stochastic}) have cut integration steps and improved quality. The framework of \citet{albergo2023stochastic} unifies flow and diffusion by bridging two densities in finite time through a time-indexed interpolant, recovering rectified flow \citep{liu2022} as a special case. Applying the generative process in the latent space of a pretrained autoencoder \citep{rombach2022high,dao2023flowmatchinglatentspace} scales these models to high-resolution signals.

These methods are commonly applied to scientific simulation.
Several works use diffusion to stabilise or correct autoregressive rollouts of spatiotemporal dynamics \citep{lippe2024pde,kohl2023turbulent,ruhling2024dyffusion} and to quantify uncertainty in under-resolved flow surrogates \citep{liu2024uncertainty}.
Diffusion has also been used as the generative core of probabilistic ensembles for medium-range weather forecasting \citep{price2023gencast} and as a posterior sampler for score-based data assimilation \citep{rozet2023scorebased}.
Most similar to our work, \citet{lino2025learning,lienen2024zero,gao2024bayesian} use diffusion models to learn the stationary distribution of developed fluid flows from transient Large Eddy or unsteady RANS simulations.
The main difference is that our work does not make any assumption on ground-truth obtained by reduced-order models and extends to a domain where no closures are available, hence all turbulent scales are resolved.



\section{Method}
\label{sec:method}
\ourmethod{} is a two-stage latent generative model. A Swin5D autoencoder first compresses 5D snapshots of the saturated distribution function into a low-dimensional latent space. A Diffusion Transformer (DiT) \citep{peebles_dit_2023} is then trained with rectified flow matching to transport a Gaussian prior to the encoded data manifold, conditioned on the operating parameters. The two stages are trained independently, and the autoencoder weights are frozen during DiT training.
Finally, to evaluate generative quality while respecting the distributional nature of turbulence, we propose \ourfid{} (\cref{sec:fid_pres}), a Fr\'echet distance in the latent space of a pretrained gyrokinetic surrogate.

\subsection{Problem setup}
\label{sec:params}
{\begingroup
\let\tempRLT\paramRLTbox
\let\tempRLn\paramRLnbox
\let\tempQ\paramQbox
\let\tempShat\paramShatbox

\renewcommand{\paramRLTbox}{\smash{\tempRLT}}
\renewcommand{\paramRLnbox}{\smash{\tempRLn}}
\renewcommand{\paramQbox}{\smash{\tempQ}}
\renewcommand{\paramShatbox}{\smash{\tempShat}}

A gyrokinetic simulation is parameterised by dimensionless numbers entering \cref{eq:gyrovlasov} through the equilibrium Maxwellian $F_{M}$ (\cref{eq:maxwellian}) and the magnetic geometry. 
\ourmethod{} conditions on four operating parameters that strongly affect ITG turbulence.
\paramRLTbox{} (normalised ion-temperature gradient) and \paramRLnbox{} (normalised density gradient) set the free-energy source, with larger $R/L_T$ pushing the simulation into strongly driven regimes and $R/L_n$ modulating the contribution of density-gradient-driven modes.
\paramQbox{} (safety factor) and \paramShatbox{} (magnetic shear) enter the magnetic geometry through the parallel dynamics and the radial wavenumber along the field line. Jointly, $(R/L_T,\: R/L_n,\: q,\: \hat{s})$ specify the stationary distribution, independently of the initial condition, and are stacked into the conditioning vector
\par\endgroup}

\begin{equation}
\label{eq:cond_vec}
\Bc = (\paramRLTbox,\;\paramRLnbox,\;\paramShatbox,\;\paramQbox).
\end{equation}

Let $f \in \mathbb{R}^{2 \times v_\parallel \times \mu \times s \times x \times y}$ denote a saturated snapshot, with the two channels carrying the real and imaginary parts of the complex distribution function. 
A velocity network $v_{\theta}(z_t, t, \Bc)$ is trained to transport $z_0\sim\mathcal{N}(0, I)$ to $z_1 = E_{\psi}(f)$ along an interpolation parameter $t \in [0,1]$, with no information about the physical time of \cref{eq:gyrovlasov}. At inference, $\hat f = D_{\psi}(z_1)$ recovers the full 5D state from the generated latent.

\subsection{Swin5D autoencoder}
$E_\psi$ and $D_\psi$ share hierarchical Swin5D layers, with n-dimensional Swin blocks functionally equivalent to GyroSwin \citep{paischer2025gyroswin}. Attention is applied to non-overlapping 5D windows of size $M = M_{v_\parallel}\!\times\!M_{\mu}\!\times\!M_{s}\!\times\!M_{x}\!\times\!M_{y}$, alternating with shifted partitions (half-window cyclic shift) to recover approximate global attention at linear cost.
The bottleneck is implemented via two Transformer layers \citep{vaswani_attention_2017} at the coarsest resolution after a linear projection to channel width $d_z$.
The operating parameters $\Bc$ are not injected into either the encoder or the decoder to decouple compression from conditioning.
Training uses a complex MSE reconstruction loss on $f$.


\subsection{Latent diffusion model}
We adhere to the stochastic-interpolant framework of \citet{albergo2023stochastic} as a rectified flow \citep{liu2022}. The prior is Gaussian, the data density is the latent representation of the saturated gyrokinetic distribution obtained via the pre-trained autoencoder, and the interpolant is the linear path of \cref{eq:fm_path}. A full derivation is given in \cref{app:generative} in the appendix.

\textbf{Flow-matching formulation}
Let $z_{1} = E_{\psi}(f)$ denote a data latent and $z_{0}\sim\mathcal{N}(0, I)$ a prior sample.
For $t\in[0,1]$, we define the linear probability path
\begin{equation}
\label{eq:fm_path}
z_{t} \;=\; t\,z_{1} + (1-t)\,z_{0},
\qquad
u^{\star}(z_{t}\mid z_{0}, z_{1}) \;=\; z_{1} - z_{0},
\end{equation}
so that the target velocity is independent of $t$.
We depart from the standard recipe in two respects to reduce path variance and linearize inference trajectories.
First, instead of independent pairings, in-batch pairs are drawn from an optimal-transport coupling $\pi(z_0, z_1)$ that minimizes the expected path length between the noise and data distributions \citep{tong2024improvinggeneralizingflowbasedgenerative}.
Second, the integration time $t$ is sampled from a logit-normal distribution $p_{\mathrm{LN}}(t)$ \citep{esser2024scalingrectifiedflowtransformers}.
The network $v_{\theta}(z_{t}, t, \Bc)$ is trained to regress $u^{\star}$ by minimizing
\begin{equation}
\label{eq:fm_loss}
\mathcal{L}_{\mathrm{FM}}(\theta) \;=\; \mathbb{E}_{\substack{z_{0}, z_{1} \sim \pi, \: t \sim p_{\mathrm{LN}}, \: \Bc}}\left\|\, v_{\theta}(z_{t}, t, \Bc) - u^\ast\,\right\|^{2}.
\end{equation}
At inference, we integrate $\dot z = v_{\theta}(z, t, \Bc)$ from $z \sim \mathcal{N}(0, I)$ with an explicit Euler scheme on a uniform grid of $N$ steps in $[0, 1]$. We use $N=15$ throughout, chosen from the accuracy-runtime sweep in \cref{app:euler_steps}.


\textbfp{Diffusion Transformer}
The velocity network $v_{\theta}$ is a DiT \citep{peebles_dit_2023} that operates directly on the autoencoder bottleneck tokens, without further patching.
The flow integration time $t$ and the four operating parameters in $\Bc$ (\cref{eq:cond_vec}) are each embedded via a sinusoidal expansion and mapped through a shared MLP to produce a joint conditioning vector.
This vector modulates each transformer block via adaptive layer normalization (adaLN) applied prior to the self-attention and feed-forward layers.
The residual gate is zero-initialised so DiT begins training as identity.

\subsection{Implementation details}
Prior work on compression \citep{galletti2026pinc} observes that training on 5D data presents instabilities, linked to the large dynamic range across velocity shells and the high dimensionality of the fields.
We introduce three interventions to stabilize model training.

\textbfp{Magnetic moment normalization}
The magnetic moment coordinate $\mu$ indexes energy shells whose amplitudes span several orders of magnitude.
A global z-score collapses small-$\mu$ contributions, whose amplitudes are orders of magnitude smaller than dominant shells.
Therefore, instead of normalizing across channels as in \citet{paischer2025gyroswin} we normalize over $\mu$ via $\bar f_{\mu}, \sigma_{f_\mu}$ over all other axes.
The resulting input to the model is then computed as $\tilde f(\cdot,\mu,\cdots) = (f - \bar f_{\mu})/(\sigma_{f_\mu} + \varepsilon)$.

\textbfp{Query-Key normalization}
Attention logits on 5D grids are prone to scale drift as the model size grows.
Inspired by the language modelling community \citep{henry2020querykey,chameleonteam2025chameleon}, RMS normalization is applied independently to the query and key tensors along the head dimension before the dot product, so that logit magnitudes are controlled independently of the feature norm.

\textbfp{Gated attention}
Also following findings in \citet{qiu2025gated}, we apply a head-wise sigmoid gate to the attention output prior to the output projection.
Let $o_{h}\in\mathbb{R}^{d_{h}}$ be the raw attention output of head $h$ and $q_{h}$ its query. We replace $o_{h}$ with $\tilde o_{h} = \sigma\!\big(W_{g}\,\mathrm{ReLU}(q_{h})\big)\odot o_{h}$,
where $W_{g}\in\mathbb{R}^{d_{h}\times d_{h}}$ is a per-head learnable projection and $\sigma (\cdot)$ the sigmoid.
The gate modulates each head contribution conditionally on its query. \citet{qiu2025gated} report that this stabilizes attention logits and yields per-head sparsity.

\subsection{Fr\'echet GyroSwin Distance (\ourfid)}
\label{sec:fid_pres}
In vision it is common to evaluate generative models at a distributional level using Fr\'echet distance based on the latent space of pre-trained classifiers \citep{Heusel2017FID,szegedy2016inception}.
In our case, there is no pre-trained classifier for gyrokinetics on which we can base our evaluation.
Furthermore, the latent space should ideally be sensitive to downstream quantities.
To this end, we extract features $h = g_{\xi}(f, \Bc)$ from a frozen and pre-trained encoder of GyroSwin \citep{paischer2025gyroswin}, fit Gaussians to generated and reference activations, and report
\begin{equation}
\label{eq:gyroswin_fid}
\text{\ourfid} \;=\; \|\mu_{r} - \mu_{g}\|_{2}^{2} \;+\; \operatorname{Tr}\!\Big(\Sigma_{r} + \Sigma_{g} - 2\,\big(\Sigma_{r}\,\Sigma_{g}\big)^{1/2}\Big).
\end{equation}
We use the publicly available checkpoint from the huggingface hub \citep{wolf2020transformers}\footnote{Checkpoints available at \url{https://huggingface.co/ml-jku/gyroswin_large}}.
GyroSwin is a UNet trained in a multitask manner to predict the time evolution of the 5D phase space, the 3D electrostatic potential fields, and the scalar heat flux.
Therefore, it provides several choices for latents that could be used for computing the \ourfid{} metric.
Also, it allows us to evaluate on latents that encode different derived physical quantities.
To provide a comprehensive evaluation, we select three different latents (bottleneck skip $L{=}2$, $\phi$ decoder, $Q$ head) and report \ourfid{} on each of them.
For further details see \cref{app:fid}.


\section{Experiments}
\label{sec:experiments}
We evaluate \ourmethod{} model along three complementary axes:
\begin{enumerate*}[label=\textbf{(\roman*)}]
    \item downstream, time-averaged physical observables and parameter scans (\cref{tab:downstream} and \cref{fig:param_scans});
    \item distributional comparisons in a pretrained latent space (\ourfid{}, \cref{tab:warmstart_fid});
    \item the convergence of the gyaradax solver \citep{galletti2026gyaradax}, a GPU accelerated version of GKW \citep{PEETERS_GKW_2009}, warm-started from generated candidates (\cref{tab:warmstart_fid}).
\end{enumerate*}
Across all evaluations, \ourmethod{} outperforms autoregressive GyroSwin and other generative baselines.

\subsection{Setup}
We evaluate \ourmethod{} on the dataset of \citet{paischer2025gyroswin}, ${\sim}250$ saturated GKW trajectories in the 4D parameter space (\cref{app:data} in appendix).\footnote{Quantized dataset at \url{https://huggingface.co/datasets/gerkone/cbc-gyroswin-256traj}.}
Three trajectories form the validation split, six the in-distribution (ID) test split, and five the out-of-distribution (OOD) split, obtained by taking points outside the training convex hull.

Three baseline families are considered.
Reduced-order models map linear growth rates to nonlinear saturated fluxes via saturation rules \citep[QuaLiKiz]{Bourdelle2015}.
Tabular regressors \citep[GPR]{Hornsby2024}) map $\Bc$ to saturated transport scalars without.
Both neglect the nonlinear physics.
Autoregressive surrogates such as \textit{GyroSwin}~\citep{paischer2025gyroswin} evolve the 5D phase space forward in time. 
We train and evaluate two GyroSwin variants.
$\text{GyroSwin (warm)}$ begins the autoregressive rollout from a semi-saturated snapshot, while $\text{GyroSwin (cold)}$ starts from an early snapshot in the linear phase (details in \cref{app:gyroswin_cold_vs_warm} in appendix).
Finally, we include three generative baselines following the same turbulence sampling perspective as \ourmethod{}. 
A VAE \citep{kingma2014vae} that decodes $z \sim \mathcal{N}(0, I)$, and two variants of a VQ-VAE \citep{vandenoord2017vqvae} that differ in how code indices are drawn at inference. 
\textit{VQ-VAE (rand)} samples them i.i.d.\ from the empirical codebook distribution, while \textit{VQ-VAE (AR)} draws them from a Transformer prior trained autoregressively over the code sequences \citep{yan2021videogptvideogenerationusing}.
See \cref{app:training} in appendix for details.

\textbfp{Latent probing}
A linear probe $\pi_\eta$ is a linear regression model fit on \ourmethod{} latents. 
It maps directly to $\bar Q$, $W(k_y)$ and $Q(k_y)$, bypassing both $D_\psi$ and the integrals from 5D to integrated quantities (\cref{eq:integrals}).
Concretely, we first apply PCA then perform Ridge regression on a training-set of (latents, $\bar Q$/$W(k_y)$/$Q(k_y)$) pairs. 
We call this method \ourmethod{} (probe), which trades accuracy for lower cost compared to \ourmethod{} (decode).


\subsection{Downstream physics accuracy}
\label{sec:downstream}
We evaluate \ourmethod{} on three time-averaged transport quantities: $\bar Q$, the binormal potential spectrum $W(k_y)$, and the flux spectrum $Q(k_y)$. The latent nature of our method admits two routes, either decoding $\hat f = D_\psi(z_1)$ or probing the latent directly. \cref{tab:downstream} includes both, but unless reported differently we decode to the 5D space.
The table is grouped in three parts, with reduced-order models at the top, autoregressive surrogates in the middle, and generative approaches at the bottom.

\ourmethod{} (probe) and \ourmethod{} (decode) achieve the lowest $\bar{Q}_\text{RMSE}$ on both ID and OOD splits, reducing the $\bar{Q}_\text{RMSE}$ of GyroSwin at a significantly lower inference cost and most importantly without the need for a semi-saturated, ``warm'' initial condition. It is also worth noting that all non-generative baselines were either solely trained on time-averaged heat flux (reduced-order models) or explicitly finetuned on it (GyroSwin flux head).
Generative baselines rely on the field integrals (\cref{sec:integrals} in appendix) to produce all derived quantities.

On the spectra, the probed latent reaches the highest $W(k_y)$ Pearson correlation on both splits. Conversely, \ourmethod{} (decode) is on-par with other methods on $W(k_y)$, while performing the best on the in-distribution $Q(k_y)$ correlation.
Quasilinear performs poorly on $W(k_y)$ as it cannot capture the required turbulent nonlinear interactions.
Since GPR directly maps to the nonlinear heat fluxes, it cannot be evaluated on the spectra.
We also report Pointwise RMSE and Wasserstein-distance in \cref{tab:downstream_spectral} in \cref{app:fid_spectral} in the appendix.
The reduced-order baselines and the latent generative baselines (VAE, VQ-VAE) consistently perform worse on both the time-averaged heat flux as well as the spectra.

\begin{table}[t]
\centering
\caption{Downstream accuracy on in-distribution (\textbf{ID}) and out-of-distribution (\textbf{OOD}) operating conditions. \textbf{5D}: produces 5D nonlinear phase space outputs. We report the RMSE of the time-averaged heat flux ($\bar Q$), Pearson correlation on $W(k_y)$, Pearson correlation on $Q(k_y)$, and wall-clock per $64$ saturated-phase samples on a single NVIDIA~H100. \ourmethod{} (probe) uses latent-space probing without 5D decoding; \ourmethod{} (decode) uses the full decode path. 
Best \textbf{bold}, second \underline{underlined}.}
\vspace{1em}
\label{tab:downstream}
\setlength{\tabcolsep}{5pt}
\renewcommand{\arraystretch}{1.1}
\small
\resizebox{\textwidth}{!}{
\begin{tabular}{l c cc cc cc c}
\toprule
\multirow{2}{*}{\textbf{Method}} & \multirow{2}{*}{\textbf{5D}}
 & \multicolumn{2}{c}{$\bar{Q}_\text{RMSE}\!\downarrow$}
 & \multicolumn{2}{c}{$W(k_y)_\text{PC}\!\uparrow$}
 & \multicolumn{2}{c}{$Q(k_y)_\text{PC}\!\uparrow$}
 & \multirow{2}{*}{\makecell{\textbf{Runtime}\,$\downarrow$\\{[in ms]}}} \\
\cmidrule(lr){3-4} \cmidrule(lr){5-6} \cmidrule(lr){7-8}
 & & \textbf{ID} & \textbf{OOD} & \textbf{ID} & \textbf{OOD} & \textbf{ID} & \textbf{OOD} & \\
\midrule
Quasilinear                   & \xmark & $56.68_{\pm 14.09}$                              & $56.06_{\pm 14.35}$                              & $0.0346_{\pm 0.1075}$                                  &     $0.0837_{\pm0.1124}$                      & $0.5260_{\pm 0.3757}$                               &     $0.5826_{\pm 0.3331}$                        & $<\!0.1$ \\
GPR                           & \xmark & $36.95_{\pm 9.24}$                              & $25.89_{\pm 8.78}$                              & ---                                  & ---                                  & ---                               & ---                               & $<\!0.1$ \\
\midrule
$\text{GyroSwin (warm)}$ & \cmark & $18.35_{\pm 13.58}$                & $26.43_{\pm 21.09}$                & $0.9689_{\pm 0.0495}$                & $0.9737_{\pm 0.0167}$                & $0.4774_{\pm 0.3097}$             & $0.6133_{\pm 0.3200}$             & $985.6$ \\
GyroSwin (cold)               & \cmark & $30.37_{\pm 19.11}$                & $33.15_{\pm 22.66}$                & $0.9671_{\pm 0.0532}$                & $0.9766_{\pm 0.0134}$                & $0.2144_{\pm 0.2188}$             & $0.2992_{\pm 0.3113}$             & $20036.3$ \\
\midrule
VAE                           & \cmark & $106.30_{\pm 32.30}$               & $112.80_{\pm 44.07}$               & $0.9663_{\pm 0.0542}$                & $0.9728_{\pm 0.0169}$                & $0.4753_{\pm 0.1921}$             & $0.4994_{\pm 0.2427}$             & \underline{18.9} \\
VQ-VAE (rand)                 & \cmark & $74.78_{\pm 29.96}$                & $72.47_{\pm 32.25}$                & $0.9687_{\pm 0.0503}$                & $0.9742_{\pm 0.0166}$                & $0.5297_{\pm 0.1826}$             & $0.5999_{\pm 0.1878}$             & 21.9 \\
VQ-VAE (AR)                   & \cmark & $52.99_{\pm 24.12}$                & $45.31_{\pm 22.82}$                & $\underline{0.9748}_{\pm 0.0407}$    & $\underline{0.9781}_{\pm 0.0172}$    & $\underline{0.8124}_{\pm 0.1850}$ & $\mathbf{0.9112}_{\pm 0.0685}$    & 132.6 \\
\textbf{\ourmethod} (probe)   & \xmark & $\underline{14.40}_{\pm 10.31}$    & $\underline{17.43}_{\pm 11.80}$    & $\mathbf{0.9812}_{\pm 0.0320}$       & $\mathbf{0.9815}_{\pm 0.0186}$       & $0.5916_{\pm 0.2330}$             & $0.6845_{\pm 0.2427}$             & \textbf{18.5} \\
\textbf{\ourmethod} (decode)  & \cmark & $\mathbf{14.36}_{\pm 9.41}$        & $\mathbf{16.38}_{\pm 7.06}$        & $0.9697_{\pm 0.0506}$                & $0.9747_{\pm 0.0175}$                & $\mathbf{0.9738}_{\pm 0.0172}$    & $\underline{0.8753}_{\pm 0.1001}$ & 35.6 \\
\bottomrule
\end{tabular}
}
\end{table}


\textbfp{Single-parameter sensitivity scans}
We investigate whether the model has learned a smooth and physically consistent attractor manifold.
To that end, we sweep each component of $\Bc$ independently across a wide ID and OOD support, with the other three fixed.
See \cref{app:scan} in appendix for details.
This generalization is non-trivial as the model is being pushed far outside of its training domain. 

\begin{figure}[t]
\centering
\includegraphics[width=\linewidth]{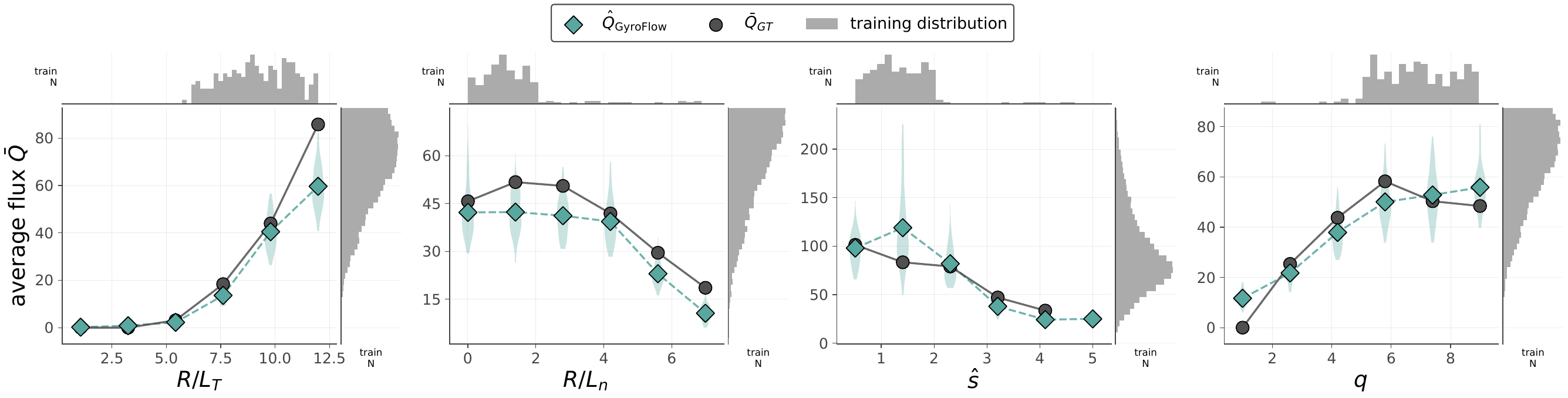}
\caption{Single-param $\bar Q$ scans along each of the four conditioning axes. Green $\diamond$: mean of \ourmethod{} samples, violins represent distribution of $K{=}64$ draws. Dark $\circ$: reference. Dark shade: training distribution over flux and each parameter.}
\label{fig:param_scans}
\end{figure}

\cref{fig:param_scans} shows that \ourmethod{} reproduces the expected ITG thresholding behavior even in regimes absent or underrepresented in the dataset.
This dataset is heavily skewed toward high $R/L_T$ ranges, and thus highly turbulent regions. \cref{fig:param_scans} includes the training distribution across parameters and flux in gray, highlighting the gaps in training data and the robustness of our method.
Surprisingly, \ourmethod{} even captures unseen near-zero fluxes at low $R/L_T$ which are entirely unseen during training.
This finding is consistent with generalization behavior for other conditioning parameters.
%

\textbfp{Runtime} \Cref{tab:downstream} reports wall-clock time on a single NVIDIA~H100 to produce $64$ samples. For $\text{GyroSwin (warm)}$ we roll forward for $64$ saturated-phase steps. We exclude the ${\sim}25$-$30$\,min GKW spin-up required to reach saturation from the reported runtime.
GyroSwin (cold) instead times the full rollout from a random perturbation through the transient plus $64$ steps, and ends up in an entirely different regime as it needs many more autoregressive calls. Per-call cost of QuaLiKiz and GPR is negligible relative to all other methods. 
\ourmethod{} attains the best performance but is orders of magnitude faster than autoregressive surrogates. 

\subsection{Distributional quality and warm-start}
\label{sec:fid_and_warmstart}

\begin{table}[b]
\centering
\caption{Distributional generation quality from both \textbf{(a)} warm-start Kolmogorov-Smirnov $\overline{D}_{S}$ and \textbf{(b)} \ourfid{} at three GyroSwin U-Net depths. Lower is better, best in \textbf{bold} and second \underline{underlined}.}
\vspace{1em}
\label{tab:warmstart_fid}
\small
\hspace{-3.5em}
\begin{minipage}[t]{0.48\textwidth}
\centering
\vspace{0pt}%
\setlength{\tabcolsep}{5pt}
\renewcommand{\arraystretch}{1.1}
\begin{tabular}{l cc}
\toprule
Method & $\overline{D}_{W(k_y)}\!\downarrow$ & $\overline{D}_{Q(k_y)}\!\downarrow$ \\
\midrule
Nearest $\Bc$ & \underline{0.363} & \underline{0.267} \\
\midrule
VAE & 0.659 & 0.471 \\
VQ-VAE (rand) & 0.642 & 0.513 \\
VQ-VAE (AR) & 0.688 & 0.497 \\
\textbf{\ourmethod{}} & \textbf{0.349} & \textbf{0.248} \\
\bottomrule
\end{tabular}
\subcaption{Warm-start KS statistic $\overline{D}_{S}$.}
\label{tab:warmstart_ad_panel}
\end{minipage}\hspace{-0.5em}
\begin{minipage}[t]{0.48\textwidth}
\centering
\vspace{0pt}%
\setlength{\tabcolsep}{6pt}
\renewcommand{\arraystretch}{1.1}
\begin{tabular}{l ccc}
\toprule
Method & skip $L{=}2$ & flux head & $\phi$ decoder \\
\midrule
VAE & 5.3e+03 & 99.3 & \underline{1.7e+05} \\
VQ-VAE (rand) & 4.6e+03 & 105.0 & 1.8e+05 \\
VQ-VAE (AR) & \underline{2.5e+03} & \underline{66.9} & \textbf{1.3e+05} \\
\ourmethod{} & \textbf{2.4e+03} & \textbf{52.5} & 3.2e+05 \\
\bottomrule
\end{tabular}
\subcaption{\ourfid{} per method at the three depths.}
\label{tab:fid_panel}
\end{minipage}
\end{table}

Since draws from \ourmethod{} are i.i.d.\ samples of the saturated distribution rather than predictions of a specific reference, evaluating generative quality requires a distributional metric.
We compare two complementary views, \begin{enumerate*}[label=(\roman*)]
    \item warm-start dynamics in physical space (the solver is initialized from $\hat f$ and we measure how close the resulting stationary marginal is to the reference) and
    \item \ourfid{}, introduced in \cref{sec:fid_pres}.
\end{enumerate*}
We find that both are cross-correlated as well as correlated to the downstream metrics.

\textbfp{Warm-starts and similarity to attractor}
A generated $\hat f$ can replace the random initial condition of numerical solvers. If $\hat f$ already lies close to the attractor $\mathbb{P}_{\Bc}$, the linear-to-nonlinear transient is shortened or completely removed.
To quantify this, for each held-out $\Bc$ we draw $R{=}5$ independent generations from \ourmethod{}, integrate each with gyaradax for $T \ll T_{\mathrm{GT}}$ steps and pair all $R$ warm-started rollouts ($\mathcal{S}^{w}_{\Bc}$) with the corresponding tail of a fully saturated reference run ($\mathcal{S}^{r}_{\Bc}$).
At every binormal mode index $\ell \in \{1, \ldots, n_{k_y}\}$, the spectrum component $S_\ell(t)$ is a scalar time series (over the rolled, warm started integration).
We compare the empirical distribution functions (CDFs) $\hat F^{w}_\ell, \hat F^{r}_\ell$ with the non-parametric two-sample Kolmogorov-Smirnov (KS) statistic \citep{Massey1951} and mean across modes:
\begin{equation*}
D_\ell \;=\; \sup_{x \in \mathbb{R}}\,\big|\hat F^{w}_\ell(x) - \hat F^{r}_\ell(x)\big| \;\in\; [0, 1],
\qquad
\overline{D}_{S}(\Bc) \;=\; \frac{1}{n_{k_y}}\sum_{\ell=1}^{n_{k_y}} D_\ell.
\end{equation*}
$D_\ell = 0$ for identical empirical CDFs and $D_\ell = 1$ for disjoint ones. The per-mode evaluation mitigates the bias introduced by the very high magnitude low-$k_y$ range ($\sim2-3$ orders of magnitude larger than the high-$k_y$). 
Details are provided in \cref{app:ks} in the appendix.
We include the nearest neighbor in the training set according to operating parameters in the comparison to the warm-restart.
The results are shown in \cref{tab:warmstart_fid}\subref{tab:warmstart_ad_panel}.
\ourmethod{} attains the lowest $\overline{D}_{S}$ on both spectra  across generative methods. 
%

\textbfp{Generation quality}
For each held-out $\Bc$ we draw $K=64$ samples $\{\hat f^{(k)}\}$ and extract features $h^{(k)} = g_{\xi}(\hat f^{(k)}, \Bc)$ from the frozen GyroSwin encoder, with the reference set produced by passing ground-truth samples at the same $\Bc$ through $g_{\xi}$. \cref{tab:warmstart_fid}\subref{tab:fid_panel} reports \ourfid{} at three U-Net depths, namely
the deepest skip $L{=}2$ (coarsest 5D resolution level of GyroSwin), the flux head activation (before the final scalar flux projection), and the $\phi$ decoder (electrostatic potential reconstruction branch).

\begin{wrapfigure}{r}{0.54\textwidth}
\vspace{-1em}
\centering
\includegraphics[width=\linewidth]{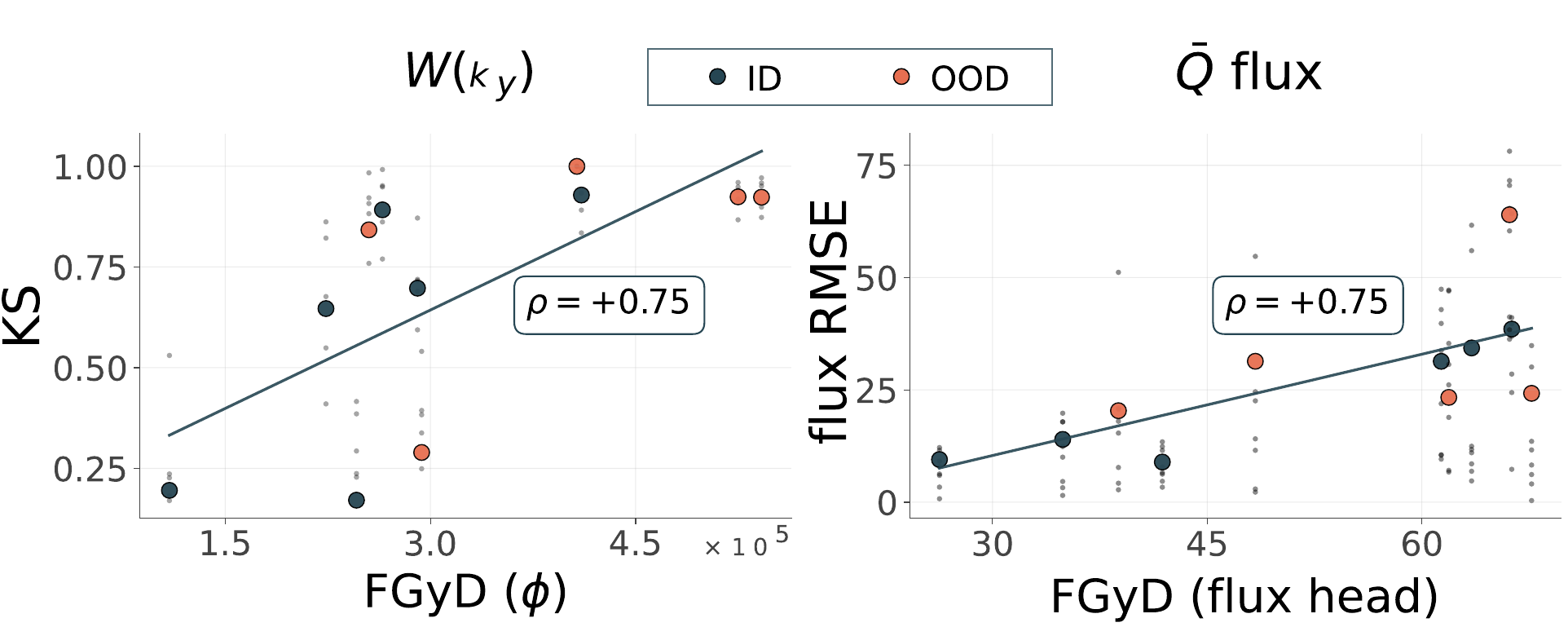}
\caption{Correlation between \ourfid{} and warm-start KS (left) and flux RMSE (right).}
\label{fig:fid_ad_corr}
\vspace{-0.6em}
\end{wrapfigure}

\ourmethod{} attains the best \ourfid{} at skip $L{=}2$ and the flux head, in line with its lead on $\bar{Q}$ and $Q(k_y)$ in \cref{tab:downstream}, indicating strong 5D generation quality.
It is worse at the $\phi$ decoder, matching the weaker $W(k_y)$ spectra reconstruction (\cref{tab:downstream}) which is itself a $\phi$-derived quantity. 
The quantitative alignment between \ourfid{} and the corresponding physical observable supports its use as a quantitative diagnostic.

To further drive this point and as a verification of both distributional metrics, we cross-validate them by pairing $\overline{D}_{S}$ and \ourfid{} per held-out $\Bc$. \cref{fig:fid_ad_corr} (left) reports a positive correlation, suggesting that latent-space distance in the GyroSwin features is linked to attractor proximity of the generated warm start candidates. A similar pattern is observed when pairing \ourfid{} on the flux head to the time-averaged flux RMSE (\cref{fig:fid_ad_corr}, right).


\section{Discussion and Conclusion}
\label{sec:discussion}

We introduced \ourmethod{}, a conditional latent flow-matching model that bypasses the transient of gyrokinetic simulations by leveraging the ergodicity assumption.
Therefore it can directly sample from the statistical 5D steady state the simulation naturally converges to.
On the cyclone-base case ITG benchmark, \ourmethod{} sets a new state of the art for heat flux prediction while providing a speedup of an order of magnitude.
%
Linear probing on the diffused latent space attains comparable accuracy while providing another two-fold speedup.
Furthermore, we demonstrate that \ourmethod{} can extrapolate along parameter regions it has not been trained on, indicating that it has learned to successfully traverse the latent attractor manifold.
Finally, we propose \ourfid{}, a distributional metric computed in the latent space of a pretrained gyrokinetic surrogate.
By warm-starting the numerical solver with generated snapshots, we show that \ourfid{} correlates with the spectral divergence of the resulting rollouts, validating it as a solver-free proxy for generation quality.

\textbfp{Limitations and future directions}
The current dataset is electrostatic, single-species, and local flux-tube.
The most direct line of follow-up is to scale \ourmethod{} to other physical regimes, namely electromagnetic fluctuations at low and high $\beta$, collisionality, and heterogeneous solvers (e.g.\ CGYRO \cite{CANDY201673}, GENE \citep{jenko2000electron}).
This requires both a heterogeneous training corpus and conditioning on additional dimensionless parameters.
A second axis is integration with transport modeling frameworks. Pairing \ourmethod{} with JINTRAC \citep{aJINTRAC-Romanelli}, TORAX \citep{citrin2024toraxfastdifferentiabletokamak} or PORTALS \citep{portals} would replace tens of thousands of numerical solver calls while ensuring significantly better accuracy than the currently used reduced-order alternatives \citep{Hornsby2024,zanisi2025,citrin2024toraxfastdifferentiabletokamak}, taking a step toward real-time transport prediction in reactor design loops.
%

\begin{ack}
This work has been funded by the Fusion Futures Programme. As announced by the UK Government in October 2023, Fusion Futures aims to provide holistic support for the development of the fusion sector.
The ELLIS Unit Linz, the LIT AI Lab, the Institute for Machine Learning, are supported by the Federal State Upper Austria. We thank the projects FWF AIRI FG 9-N (10.55776/FG9), AI4GreenHeatingGrids (FFG- 899943), Stars4Waters (HORIZON-CL6-2021-CLIMATE-01-01), FWF Bilateral Artificial Intelligence (10.55776/COE12). We thank NXAI GmbH, Audi AG, Merck Healthcare KGaA, GLS (Univ. Waterloo), T\"{U}V Holding GmbH, Software Competence Center Hagenberg GmbH, dSPACE GmbH, TRUMPF SE + Co. KG.
We acknowledge EuroHPC Joint Undertaking for awarding us access to Leonardo at CINECA, Italy, and Deucalion at MACC, Portugal.
The authors acknowledge the use of resources provided by the Isambard-AI National AI Research Resource (AIRR). Isambard-AI is operated by the University of Bristol and is funded by the UK Government's Department for Science, Innovation and Technology (DSIT) via UK Research and Innovation; and the Science and Technology Facilities Council [ST/AIRR/I-A-I/1023].
\end{ack}

\bibliographystyle{neurips_2025}
\bibliography{references}

@inproceedings{zanisi2025,
  author       = {Zanisi, Lorenzo and Järvinen, Aaro and Kit, Adam and Bruncrona, Amanda and Fernando, Anushan and Patel, Bhavin and Siddle, Catherine and Roach, Colin and Jordan, Daniel and Casson, Francis and Eriksson, Frida and Snoep, Garud and Dudding, Harry and Buchanan, James and Citrin, Jonathan and Quarantiello, Luigi and Meneghini, Orso and Hamel, P. and Correnti, Salvatore and Pamela, Stanislas and Norman, T. and Brown, Theodore and Neiser, Tom and Lomonaco, Vincenzo},
  title        = {Data-Efficient Digital Twinning Strategies and Surrogate Models of Quasilinear Turbulence in {JET} and {STEP}},
  booktitle    = {Proceedings of the 30th IAEA Fusion Energy Conference (FEC 2025)},
  year         = {2025},
  month        = oct,
  address      = {Chengdu, China},
  organization = {International Atomic Energy Agency (IAEA)},
  note         = {Paper ID: TH-C, Contribution 36059}
}

@article{Bourdelle2025,
  title = {Integrated modelling of tokamak plasmas: progress and challenges towards ITER operation and reactor design},
  volume = {67},
  ISSN = {1361-6587},
  DOI = {10.1088/1361-6587/adc484},
  number = {4},
  journal = {Plasma Physics and Controlled Fusion},
  publisher = {IOP Publishing},
  author = {Bourdelle,  C},
  year = {2025},
  month = apr,
  pages = {043001}
}

@article{Bourdelle2015,
  doi = {10.1088/0741-3335/58/1/014036},
  year = {2015},
  month = dec,
  publisher = {{IOP} Publishing},
  volume = {58},
  number = {1},
  pages = {014036},
  author = {C Bourdelle and J Citrin and B Baiocchi and A Casati and P Cottier and X Garbet and F Imbeaux and},
  title = {Core turbulent transport in tokamak plasmas: bridging theory and experiment with {QuaLiKiz}},
  journal = {Plasma Physics and Controlled Fusion}
}

@article{VanMulders2021,
  doi = {10.1088/1741-4326/ac0d12},
  year = {2021},
  month = jul,
  publisher = {{IOP} Publishing},
  volume = {61},
  number = {8},
  pages = {086019},
  author = {S. Van Mulders and F. Felici and O. Sauter and J. Citrin and A. Ho and M. Marin and K.L. van de Plassche},
  title = {Rapid optimization of stationary tokamak plasmas in {RAPTOR}: demonstration for the {ITER} hybrid scenario with neural network surrogate transport model {QLKNN}},
  journal = {Nuclear Fusion}
}

@article{Citrin_qlknn_2023,
	doi = {10.1063/5.0136752},
	year = 2023,
	month = {jun},
	publisher = {{AIP} Publishing},
	volume = {30},
	number = {6},
	author = {J. Citrin and P. Trochim and T. Goerler and D. Pfau and K. L. van de Plassche and F. Jenko},
	title = {Fast transport simulations with higher-fidelity surrogate models for {ITER}},
	journal = {Physics of Plasmas}
}

@article{Hornsby2024,
  title = {Gaussian process regression models for the properties of micro-tearing modes in spherical tokamaks},
  volume = {31},
  ISSN = {1089-7674},
  DOI = {10.1063/5.0174478},
  number = {1},
  journal = {Physics of Plasmas},
  publisher = {AIP Publishing},
  author = {Hornsby,  W. A and Gray,  A. and Buchanan,  J. and Patel,  B. S. and Kennedy,  D. and Casson,  F. J. and Roach,  C. M. and Lykkegaard,  M. B. and Nguyen,  H. and Papadimas,  N. and Fourcin,  B. and Hart,  J.},
  year = {2024},
  month = {jan} 
}

@article{Citrin2017,
  doi = {10.1088/1361-6587/aa8aeb},
  year = {2017},
  month = nov,
  publisher = {{IOP} Publishing},
  volume = {59},
  number = {12},
  pages = {124005},
  author = {J Citrin and C Bourdelle and F J Casson and C Angioni and N Bonanomi and Y Camenen and X Garbet and L Garzotti and T G\"{o}rler and O G\"{u}rcan and F Koechl and F Imbeaux and O Linder and K van de Plassche and P Strand and G Szepesi and},
  title = {Tractable flux-driven temperature,  density,  and rotation profile evolution with the quasilinear gyrokinetic transport model {QuaLiKiz}},
  journal = {Plasma Physics and Controlled Fusion}
}

@article{Staebler2007,
  title = {A theory-based transport model with comprehensive physics},
  volume = {14},
  ISSN = {1089-7674},
  DOI = {10.1063/1.2436852},
  number = {5},
  journal = {Physics of Plasmas},
  publisher = {AIP Publishing},
  author = {Staebler,  G. M. and Kinsey,  J. E. and Waltz,  R. E.},
  year = {2007},
  month = {may} 
}

@article{Staebler2010,
  title = {Electron collisions in the trapped gyro-Landau fluid transport model},
  volume = {17},
  ISSN = {1089-7674},
  DOI = {10.1063/1.3505308},
  number = {12},
  journal = {Physics of Plasmas},
  publisher = {AIP Publishing},
  author = {Staebler,  G. M. and Kinsey,  J. E.},
  year = {2010},
  month = {dec}
}

@article{Zanisi2024,
  title = {Efficient training sets for surrogate models of tokamak turbulence with Active Deep Ensembles},
  volume = {64},
  ISSN = {1741-4326},
  DOI = {10.1088/1741-4326/ad240d},
  number = {3},
  journal = {Nuclear Fusion},
  publisher = {IOP Publishing},
  author = {Zanisi,  L. and Ho,  A. and Barr,  J. and Madula,  T. and Citrin,  J. and Pamela,  S. and Buchanan,  J. and Casson,  F.J. and Gopakumar,  V.},
  year = {2024},
  month = feb,
  pages = {036022}
}

@article{Kennedy2023,
  title = {Electromagnetic gyrokinetic instabilities in STEP},
  volume = {63},
  ISSN = {1741-4326},
  DOI = {10.1088/1741-4326/ad08e7},
  number = {12},
  journal = {Nuclear Fusion},
  publisher = {IOP Publishing},
  author = {Kennedy,  D. and Giacomin,  M. and Casson,  F.J. and Dickinson,  D. and Hornsby,  W.A. and Patel,  B.S. and Roach,  C.M.},
  year = {2023},
  month = {nov},
  pages = {126061}
}

@article{aJINTRAC-Romanelli,
	author = {Romanelli, M and Corrigan, G and Parail, V and Wiesen, Sven and Ambrosino, Roberto and Da Silva Aresta Belo, P and Garzotti, Luca and Harting, P and K{\"o}chl, F and Koskela, Tuomas and Lauro-Taroni, L and Marchetto, Chiara and Mattei, Massimiliano and Militello-Asp, E and Nave, M and Pamela, Stanislas and Salmi, A and Strand, P and Szepesi, G},
	year = {2014},
	month = {01},
	title = {{JINTRAC}: {A} system of codes for integrated simulation of Tokamak scenarios},
	volume = {9},
	journal = {Plasma and Fusion Research},
	doi = {10.1585/pfr.9.3403023}
}

@article{plassche_qlknn_2020,
    author = {van de Plassche, K. L. and Citrin, J. and Bourdelle, C. and Camenen, Y. and Casson, F. J. and Dagnelie, V. I. and Felici, F. and Ho, A. and Van Mulders, S. and JET Contributors},
    title = {Fast modeling of turbulent transport in fusion plasmas using neural networks},
    journal = {Physics of Plasmas},
    volume = {27},
    number = {2},
    pages = {022310},
    year = {2020},
    month = {02},
    issn = {1070-664X},
    doi = {10.1063/1.5134126},
    eprint = {https://pubs.aip.org/aip/pop/article-pdf/doi/10.1063/1.5134126/15772521/022310\_1\_online.pdf},
}

@article{narita_toward_2022,
doi = {10.1088/1741-4326/ac70e8},
year = {2022},
month = {jun},
publisher = {IOP Publishing},
volume = {62},
number = {8},
pages = {086037},
author = {Narita, E. and Honda, M. and Maeyama, S. and Watanabe, T.-H.},
title = {Toward efficient runs of nonlinear gyrokinetic simulations assisted by a convolutional neural network model recognizing wavenumber-space images},
journal = {Nuclear Fusion},
}

@article{mitsuru_multimodal_2023,
author = {Honda, Mitsuru and Narita, Emi and Maeyama, Shinya and Watanabe, Tomo-Hiko},
title = {Multimodal convolutional neural networks for predicting evolution of gyrokinetic simulations},
journal = {Contributions to Plasma Physics},
volume = {63},
number = {5-6},
pages = {e202200137},
doi = {https://doi.org/10.1002/ctpp.202200137},
eprint = {https://onlinelibrary.wiley.com/doi/pdf/10.1002/ctpp.202200137},
year = {2023}
}

@article{PEETERS_GKW_2009,
title = {The nonlinear gyro-kinetic flux tube code GKW},
journal = {Computer Physics Communications},
volume = {180},
number = {12},
pages = {2650-2672},
year = {2009},
note = {40 YEARS OF CPC: A celebratory issue focused on quality software for high performance, grid and novel computing architectures},
issn = {0010-4655},
doi = {https://doi.org/10.1016/j.cpc.2009.07.001},
author = {A.G. Peeters and Y. Camenen and F.J. Casson and W.A. Hornsby and A.P. Snodin and D. Strintzi and G. Szepesi}
}

@article{bourdelle_qualikiz_2007,
    author = {Bourdelle, C. and Garbet, X. and Imbeaux, F. and Casati, A. and Dubuit, N. and Guirlet, R. and Parisot, T.},
    title = {A new gyrokinetic quasilinear transport model applied to particle transport in tokamak plasmas},
    journal = {Physics of Plasmas},
    volume = {14},
    number = {11},
    pages = {112501},
    year = {2007},
    month = {11},
    issn = {1070-664X},
    doi = {10.1063/1.2800869},
    eprint = {https://pubs.aip.org/aip/pop/article-pdf/doi/10.1063/1.2800869/16746472/112501\_1\_online.pdf},
}

@article{Itoh_ZF_06,
    author = {Itoh, K. and Itoh, S.-I. and Diamond, P. H. and Hahm, T. S. and Fujisawa, A. and Tynan, G. R. and Yagi, M. and Nagashima, Y.},
    title = {Physics of zonal flowsa)},
    journal = {Physics of Plasmas},
    volume = {13},
    number = {5},
    pages = {055502},
    year = {2006},
    month = {05},
    issn = {1070-664X},
    doi = {10.1063/1.2178779},
    eprint = {https://pubs.aip.org/aip/pop/article-pdf/doi/10.1063/1.2178779/15800949/055502\_1\_online.pdf},
}

@article{Krommes_gyrokinetics_2012,
   author = "Krommes, John A.",
   title = "The Gyrokinetic Description of Microturbulence in Magnetized Plasmas", 
   journal= "Annual Review of Fluid Mechanics",
   year = "2012",
   volume = "44",
   number = "Volume 44, 2012",
   pages = "175-201",
   doi = "https://doi.org/10.1146/annurev-fluid-120710-101223",
   publisher = "Annual Reviews",
   issn = "1545-4479",
   type = "Journal Article",
  }

@article{Staebler_quasilinear_2024,
doi = {10.1088/1741-4326/ad6ba5},
year = {2024},
month = {sep},
publisher = {IOP Publishing},
volume = {64},
number = {10},
pages = {103001},
author = {Staebler, G. and Bourdelle, C. and Citrin, J. and Waltz, R.},
title = {Quasilinear theory and modelling of gyrokinetic turbulent transport in tokamaks},
journal = {Nuclear Fusion}
}

@article{Kiefer_quasilinear_2021,
doi = {10.1088/1741-4326/abfc9c},
year = {2021},
month = {may},
publisher = {IOP Publishing},
volume = {61},
number = {6},
pages = {066035},
author = {Kiefer, C.K. and Angioni, C. and Tardini, G. and Bonanomi, N. and Geiger, B. and Mantica, P. and Pütterich, T. and Fable, E. and Schneider, P.A. and , ASDEX Upgrade Team and , EUROfusion MST1 Team and , JET Contributors},
title = {Validation of quasi-linear turbulent transport models against plasmas with dominant electron heating for the prediction of ITER PFPO-1 plasmas},
journal = {Nuclear Fusion}
}

@article{dimits_shift_2000,
    author = {Dimits, A. M. and Bateman, G. and Beer, M. A. and Cohen, B. I. and Dorland, W. and Hammett, G. W. and Kim, C. and Kinsey, J. E. and Kotschenreuther, M. and Kritz, A. H. and Lao, L. L. and Mandrekas, J. and Nevins, W. M. and Parker, S. E. and Redd, A. J. and Shumaker, D. E. and Sydora, R. and Weiland, J.},
    title = {Comparisons and physics basis of tokamak transport models and turbulence simulations},
    journal = {Physics of Plasmas},
    volume = {7},
    number = {3},
    pages = {969-983},
    year = {2000},
    month = {03},
    issn = {1070-664X},
    doi = {10.1063/1.873896},
    eprint = {https://pubs.aip.org/aip/pop/article-pdf/7/3/969/19007104/969\_1\_online.pdf},
}

@inproceedings{Bourdelle_validity_2008,
  author       = {C. Bourdelle and A. Casati and X. Garbet and F. Imbeaux and J. Candy and F. Clairet and G. Dif-Pradalier and G. Falchetto and T. Gerbaud and V. Grandgirard and P. Hennequin and R. Sabot and Y. Sarazin and L. Vermare and R. E. Waltz},
  title        = {Validity of Quasi-Linear Transport Model},
  booktitle    = {Proceedings of the 22nd IAEA Fusion Energy Conference},
  year         = {2008},
  pages        = {227},
  publisher    = {International Atomic Energy Agency},
  address      = {Vienna, Austria},
  note         = {Paper TH/P8-7}
}

@inproceedings{vaswani_attention_2017,
  author       = {Ashish Vaswani and
                  Noam Shazeer and
                  Niki Parmar and
                  Jakob Uszkoreit and
                  Llion Jones and
                  Aidan N. Gomez and
                  Lukasz Kaiser and
                  Illia Polosukhin},
  editor       = {Isabelle Guyon and
                  Ulrike von Luxburg and
                  Samy Bengio and
                  Hanna M. Wallach and
                  Rob Fergus and
                  S. V. N. Vishwanathan and
                  Roman Garnett},
  title        = {Attention is All you Need},
  booktitle    = {Advances in Neural Information Processing Systems 30: Annual Conference
                  on Neural Information Processing Systems 2017, December 4-9, 2017,
                  Long Beach, CA, {USA}},
  pages        = {5998--6008},
  year         = {2017},
  bibsource    = {dblp computer science bibliography, https://dblp.org}
}

@article{Wan_itgtransfer_2025,
doi = {10.1088/1741-4326/adc7c9},
year = {2025},
month = {apr},
publisher = {IOP Publishing},
volume = {65},
number = {5},
pages = {054001},
author = {Wan, Chenguang and Cho, Youngwoo and Qu, Zhisong and Camenen, Yann and Varennes, Robin and Lim, Kyungtak and Li, Kunpeng and Li, Jiangang and Li, Yanlong and Garbet, Xavier},
title = {A high-fidelity surrogate model for the ion temperature gradient (ITG) instability using a small expensive simulation dataset},
journal = {Nuclear Fusion}
}

@inproceedings{peebles_dit_2023,
  author       = {William Peebles and
                  Saining Xie},
  title        = {Scalable Diffusion Models with Transformers},
  booktitle    = {{IEEE/CVF} International Conference on Computer Vision, {ICCV} 2023,
                  Paris, France, October 1-6, 2023},
  pages        = {4172--4182},
  publisher    = {{IEEE}},
  year         = {2023},
  doi          = {10.1109/ICCV51070.2023.00387},
  bibsource    = {dblp computer science bibliography, https://dblp.org}
}

@misc{citrin2024toraxfastdifferentiabletokamak,
      title={TORAX: A Fast and Differentiable Tokamak Transport Simulator in JAX}, 
      author={Jonathan Citrin and Ian Goodfellow and Akhil Raju and Jeremy Chen and Jonas Degrave and Craig Donner and Federico Felici and Philippe Hamel and Andrea Huber and Dmitry Nikulin and David Pfau and Brendan Tracey and Martin Riedmiller and Pushmeet Kohli},
      year={2024},
      eprint={2406.06718},
      archivePrefix={arXiv},
      primaryClass={physics.plasm-ph},
}

@article{dimits_cbc_2000,
    author = {Dimits, A. M. and Bateman, G. and Beer, M. A. and Cohen, B. I. and Dorland, W. and Hammett, G. W. and Kim, C. and Kinsey, J. E. and Kotschenreuther, M. and Kritz, A. H. and Lao, L. L. and Mandrekas, J. and Nevins, W. M. and Parker, S. E. and Redd, A. J. and Shumaker, D. E. and Sydora, R. and Weiland, J.},
    title = {Comparisons and physics basis of tokamak transport models and turbulence simulations},
    journal = {Physics of Plasmas},
    volume = {7},
    number = {3},
    pages = {969-983},
    year = {2000},
    month = {03},
    issn = {1070-664X},
    doi = {10.1063/1.873896},
    eprint = {https://pubs.aip.org/aip/pop/article-pdf/7/3/969/19007104/969_1_online.pdf},
}

@inproceedings{paischer2025gyroswin,
title={GyroSwin: 5D Surrogates for Gyrokinetic Plasma Turbulence Simulations},
author={Fabian Paischer and Gianluca Galletti and William Hornsby and Paul Setinek and Lorenzo Zanisi and Naomi Carey and Stanislas Pamela and Johannes Brandstetter},
booktitle={The Thirty-ninth Annual Conference on Neural Information Processing Systems},
year={2025},
}

@misc{galletti2026pinc,
      title={Physics-Informed Neural Compression of High-Dimensional Plasma Data}, 
      author={Gianluca Galletti and Gerald Gutenbrunner and Sandeep S. Cranganore and William Hornsby and Lorenzo Zanisi and Naomi Carey and Stanislas Pamela and Johannes Brandstetter and Fabian Paischer},
      year={2026},
      eprint={2602.04758},
      archivePrefix={arXiv},
      primaryClass={physics.plasm-ph}
}

@misc{albergo2023stochastic,
  doi = {10.48550/ARXIV.2303.08797},
  author = {Albergo, Michael S. and Boffi, Nicholas M. and Vanden-Eijnden, Eric},
  title = {Stochastic Interpolants: A Unifying Framework for Flows and Diffusions},
  publisher = {arXiv},
  year = {2023},
}

@inproceedings{
liu2022,
title={Flow Straight and Fast: Learning to Generate and Transfer Data with Rectified Flow},
author={Xingchao Liu and Chengyue Gong and Qiang Liu},
booktitle={The Eleventh International Conference on Learning Representations },
year={2023}
}

@misc{tong2024improvinggeneralizingflowbasedgenerative,
      title={Improving and generalizing flow-based generative models with minibatch optimal transport}, 
      author={Alexander Tong and Kilian Fatras and Nikolay Malkin and Guillaume Huguet and Yanlei Zhang and Jarrid Rector-Brooks and Guy Wolf and Yoshua Bengio},
      year={2024},
      eprint={2302.00482},
      archivePrefix={arXiv},
      primaryClass={cs.LG}
}

@misc{esser2024scalingrectifiedflowtransformers,
      title={Scaling Rectified Flow Transformers for High-Resolution Image Synthesis}, 
      author={Patrick Esser and Sumith Kulal and Andreas Blattmann and Rahim Entezari and Jonas Müller and Harry Saini and Yam Levi and Dominik Lorenz and Axel Sauer and Frederic Boesel and Dustin Podell and Tim Dockhorn and Zion English and Kyle Lacey and Alex Goodwin and Yannik Marek and Robin Rombach},
      year={2024},
      eprint={2403.03206},
      archivePrefix={arXiv},
      primaryClass={cs.CV}
}

@misc{szegedy2016inception,
      title={Inception-v4, Inception-ResNet and the Impact of Residual Connections on Learning}, 
      author={Christian Szegedy and Sergey Ioffe and Vincent Vanhoucke and Alex Alemi},
      year={2016},
      eprint={1602.07261},
      archivePrefix={arXiv},
      primaryClass={cs.CV},
}

@misc{qiu2025gated,
      title={Gated Attention for Large Language Models: Non-linearity, Sparsity, and Attention-Sink-Free}, 
      author={Zihan Qiu and Zekun Wang and Bo Zheng and Zeyu Huang and Kaiyue Wen and Songlin Yang and Rui Men and Le Yu and Fei Huang and Suozhi Huang and Dayiheng Liu and Jingren Zhou and Junyang Lin},
      year={2025},
      eprint={2505.06708},
      archivePrefix={arXiv},
      primaryClass={cs.CL},
}

@misc{henry2020querykey,
      title={Query-Key Normalization for Transformers},
      author={Alex Henry and Prudhvi Raj Dachapally and Shubham Pawar and Yuxuan Chen},
      year={2020},
      eprint={2010.04245},
      archivePrefix={arXiv},
      primaryClass={cs.CL},
}

@inproceedings{lino2025learning,
      title={Learning Distributions of Complex Fluid Simulations with Diffusion Graph Networks},
      author={Mario Lino and Tobias Pfaff and Nils Thuerey},
      booktitle={International Conference on Learning Representations},
      year={2025},
}

@article{gao2024bayesian,
  title={Bayesian conditional diffusion models for versatile spatiotemporal turbulence generation},
  author={Gao, Han and Li, Luning and Jiao, Xuhui and Anandkumar, Anima},
  journal={Computer Methods in Applied Mechanics and Engineering},
  volume={427},
  pages={117023},
  year={2024},
}

@inproceedings{lienen2024zero,
  title={From Zero to Turbulence: Generative Modeling for 3D Flow Simulation},
  author={Lienen, Marten and L{\"u}dke, David and Hansen-Palmus, Jan and G{\"u}nnemann, Stephan},
  booktitle={Proceedings of the 12$^\text{th}$ International Conference on Learning Representations},
  year={2024},
}

@book{pope2000turbulent,
  title={Turbulent flows},
  author={Pope, Stephen B},
  year={2000},
  publisher={Cambridge University Press},
}

@article{lippe2024pde,
  title={{PDE}-Refiner: Achieving Accurate Long Rollouts with Neural {PDE} Solvers},
  author={Lippe, Phillip and Veeling, Bastiaan S and Perdikaris, Paris and Turner, Richard E and Brandstetter, Johannes},
  journal={Advances in Neural Information Processing Systems},
  year={2024},
}

@article{kohl2023turbulent,
  title={Turbulent flow simulation using autoregressive conditional diffusion models},
  author={Kohl, Georg and Chen, Li-Wei and Thuerey, Nils},
  journal={arXiv preprint arXiv:2309.01745},
  year={2023},
}

@techreport{slotnick2014cfd2030,
    author      = {Slotnick, Jeffrey and Khodadoust, Abdollah and Alonso, Juan
                   and Darmofal, David and Gropp, William and Lurie, Elizabeth
                   and Mavriplis, Dimitri},
    title       = {{CFD Vision 2030 Study: A Path to Revolutionary Computational
                   Aerosciences}},
    institution = {NASA Langley Research Center},
    number      = {NASA/CR-2014-218178},
    year        = {2014},
}

@misc{portals,
      title={Enhancing predictive capabilities in fusion burning plasmas through surrogate-based optimization in core transport solvers},
      author={P. Rodriguez-Fernandez and N. T. Howard and A. Saltzman and S. Kantamneni and J. Candy and C. Holland and M. Balandat and S. Ament and A. E. White},
      year={2024},
      eprint={2312.12610},
      archivePrefix={arXiv},
      primaryClass={physics.plasm-ph}
}

@article{ho2020denoising,
  title={Denoising diffusion probabilistic models},
  author={Ho, Jonathan and Jain, Ajay and Abbeel, Pieter},
  journal={Advances in neural information processing systems},
  volume={33},
  pages={6840--6851},
  year={2020}
}

@article{song2020improved,
  title={Improved techniques for training score-based generative models},
  author={Song, Yang and Ermon, Stefano},
  journal={Advances in neural information processing systems},
  volume={33},
  pages={12438--12448},
  year={2020}
}

@inproceedings{song2021denoising,
  title={Denoising Diffusion Implicit Models},
  author={Song, Jiaming and Meng, Chenlin and Ermon, Stefano},
  booktitle={Proceedings of the 9th International Conference on Learning Representations},
  year={2021}
}

@article{dhariwal2021diffusion,
  title={Diffusion models beat {GAN}s on image synthesis},
  author={Dhariwal, Prafulla and Nichol, Alexander},
  journal={Advances in neural information processing systems},
  volume={34},
  pages={8780--8794},
  year={2021}
}

@inproceedings{rombach2022high,
  title={High-resolution image synthesis with latent diffusion models},
  author={Rombach, Robin and Blattmann, Andreas and Lorenz, Dominik and Esser, Patrick and Ommer, Bj{\"o}rn},
  booktitle={Proceedings of the IEEE/CVF conference on computer vision and pattern recognition},
  pages={10684--10695},
  year={2022}
}

@inproceedings{karras2022_Elucidating,
  title={Elucidating the Design Space of Diffusion-Based Generative Models},
  booktitle={Advances in Neural Information Processing Systems 35},
  author={Karras, Tero and Aittala, Miika and Aila, Timo and Laine, Samuli},
  year={2022}
}

@inproceedings{lipmanflow,
  title={Flow Matching for Generative Modeling},
  author={Lipman, Yaron and Chen, Ricky T. Q. and Ben-Hamu, Heli and Nickel, Maximilian and Le, Matthew},
  booktitle={Proceedings of the 11th International Conference on Learning Representations},
  year={2023}
}

@article{ruhling2024dyffusion,
  title={{DYffusion}: A dynamics-informed diffusion model for spatiotemporal forecasting},
  author={R{\"u}hling Cachay, Salva and Zhao, Bo and Joren, Hailey and Yu, Rose},
  journal={Advances in Neural Information Processing Systems},
  volume={36},
  year={2024}
}

@article{liu2024uncertainty,
  title={Uncertainty-Aware Surrogate Models for Airfoil Flow Simulations with Denoising Diffusion Probabilistic Models},
  author={Liu, Qiang and Thuerey, Nils},
  journal={AIAA Journal},
  year={2024}
}

@misc{yang2025diffusionmodelscomprehensivesurvey,
      title={Diffusion Models: A Comprehensive Survey of Methods and Applications},
      author={Ling Yang and Zhilong Zhang and Yang Song and Shenda Hong and Runsheng Xu and Yue Zhao and Wentao Zhang and Bin Cui and Ming-Hsuan Yang},
      year={2025},
      eprint={2209.00796},
      archivePrefix={arXiv},
      primaryClass={cs.LG},
}

@misc{chameleonteam2025chameleon,
      title={Chameleon: Mixed-Modal Early-Fusion Foundation Models}, 
      author={Chameleon Team},
      year={2025},
      eprint={2405.09818},
      archivePrefix={arXiv},
      primaryClass={cs.CL},
}

@misc{dao2023flowmatchinglatentspace,
      title={Flow Matching in Latent Space}, 
      author={Quan Dao and Hao Phung and Binh Nguyen and Anh Tran},
      year={2023},
      eprint={2307.08698},
      archivePrefix={arXiv},
      primaryClass={cs.CV},
}

@inproceedings{perez2018film,
  title = {FiLM: Visual Reasoning with a General Conditioning Layer},
  author = {Perez, Ethan and Strub, Florian and de Vries, Harm and Dumoulin, Vincent and Courville, Aaron},
  booktitle = {AAAI},
  year = {2018},
}

@inproceedings{Heusel2017FID,
  title={GANs trained by a two time-scale update rule converge to a local Nash equilibrium},
  author={Heusel, Martin and Ramsauer, Hubert and Unterthiner, Thomas and Nessler, Bernhard and Hochreiter, Sepp},
  booktitle={Advances in Neural Information Processing Systems},
  year={2017}
}

@inproceedings{Binkowski2018DemystifyingMMD,
  title={Demystifying {MMD} {GANs}},
  author={Bi{\'n}kowski, Miko{\l}aj and Sutherland, Danica J. and Arbel, Michael and Gretton, Arthur},
  booktitle={International Conference on Learning Representations},
  year={2018}
}

@inproceedings{Chong2020UnbiasedFID,
  title={Effectively Unbiased {FID} and {Inception} Score and Where to Find Them},
  author={Chong, Min Jin and Forsyth, David},
  booktitle={IEEE Conference on Computer Vision and Pattern Recognition},
  year={2020}
}

@article{Massey1951,
  author  = {Massey, Frank J.},
  title   = {The {K}olmogorov--{S}mirnov Test for Goodness of Fit},
  journal = {Journal of the American Statistical Association},
  volume  = {46},
  number  = {253},
  pages   = {68--78},
  year    = {1951},
  publisher = {Taylor \& Francis},
  doi     = {10.1080/01621459.1951.10500769},
}

@misc{price2023gencast,
  title         = {{GenCast}: Diffusion-based ensemble forecasting for medium-range weather},
  author        = {Price, Ilan and Sanchez-Gonzalez, Alvaro and Alet, Ferran and Andersson, Tom R. and El-Kadi, Andrew and Masters, Dominic and Ewalds, Timo and Stott, Jacklynn and Mohamed, Shakir and Battaglia, Peter and Lam, Remi and Willson, Matthew},
  year          = {2023},
  eprint        = {2312.15796},
  archivePrefix = {arXiv},
  primaryClass  = {cs.LG},
}

@misc{rozet2023scorebased,
  title         = {Score-based Data Assimilation},
  author        = {Rozet, Fran\c{c}ois and Louppe, Gilles},
  year          = {2023},
  eprint        = {2306.10574},
  archivePrefix = {arXiv},
  primaryClass  = {cs.LG},
}

@article{Federrath2012,
  title         = {The Star Formation Rate of Turbulent Magnetized Clouds: Comparing Theory, Simulations, and Observations},
  author        = {Federrath, Christoph and Klessen, Ralf S.},
  journal       = {The Astrophysical Journal},
  volume        = {761},
  number        = {2},
  pages         = {156},
  year          = {2012},
  eprint        = {1209.2856},
  archivePrefix = {arXiv},
  primaryClass  = {astro-ph.SR},
  doi           = {10.1088/0004-637X/761/2/156},
}

@article{pitsch2006large,
  title={Large-eddy simulation of turbulent combustion},
  author={Pitsch, Heinz},
  journal={Annu. Rev. Fluid Mech.},
  volume={38},
  number={1},
  pages={453--482},
  year={2006},
  publisher={Annual Reviews}
}

@article{jenko2000electron,
  title={Electron temperature gradient driven turbulence},
  author={Kotschenreuther, M and Rogers, BN},
  journal={Physics of plasmas},
  volume={7},
  number={5},
  pages={1904--1910},
  year={2000},
  publisher={American Institute of Physics}
}

@article{CANDY201673,
title = {A high-accuracy Eulerian gyrokinetic solver for collisional plasmas},
journal = {Journal of Computational Physics},
volume = {324},
pages = {73-93},
year = {2016},
issn = {0021-9991},
author = {J. Candy and E.A. Belli and R.V. Bravenec},
}

@inproceedings{kingma2014vae,
  author       = {Diederik P. Kingma and
                  Max Welling},
  editor       = {Yoshua Bengio and
                  Yann LeCun},
  title        = {Auto-Encoding Variational Bayes},
  booktitle    = {2nd International Conference on Learning Representations, {ICLR} 2014,
                  Banff, AB, Canada, April 14-16, 2014, Conference Track Proceedings},
  year         = {2014}
}

@inproceedings{vandenoord2017vqvae,
 author = {van den Oord, Aaron and Vinyals, Oriol and kavukcuoglu, koray},
 booktitle = {Advances in Neural Information Processing Systems},
 editor = {I. Guyon and U. Von Luxburg and S. Bengio and H. Wallach and R. Fergus and S. Vishwanathan and R. Garnett},
 pages = {},
 publisher = {Curran Associates, Inc.},
 title = {Neural Discrete Representation Learning},
 volume = {30},
 year = {2017}
}

@inproceedings{wolf2020transformers,
  title={Transformers: State-of-the-art natural language processing},
  author={Wolf, Thomas and Debut, Lysandre and Sanh, Victor and Chaumond, Julien and Delangue, Clement and Moi, Anthony and Cistac, Pierric and Rault, Tim and Louf, R{\'e}mi and Funtowicz, Morgan and others},
  booktitle={Proceedings of the 2020 conference on empirical methods in natural language processing: system demonstrations},
  pages={38--45},
  year={2020}
}

@misc{galletti2026gyaradax,
      title={gyaradax: Local Gyrokinetics JAX Code}, 
      author={Gianluca Galletti and Eric Volkmann and Johannes Brandstetter},
      year={2026},
      eprint={2604.06085},
      archivePrefix={arXiv},
      primaryClass={physics.plasm-ph},
}

@article{coppi_1967_itg,
    author = {Coppi, B. and Rosenbluth, M. N. and Sagdeev, R. Z.},
    title = {Instabilities due to Temperature Gradients in Complex Magnetic Field Configurations},
    journal = {The Physics of Fluids},
    volume = {10},
    number = {3},
    pages = {582-587},
    year = {1967},
    month = {03},
    issn = {0031-9171},
    doi = {10.1063/1.1762151},
    eprint = {https://pubs.aip.org/aip/pfl/article-pdf/10/3/582/12589339/582_1_online.pdf},
}

@misc{yan2021videogptvideogenerationusing,
      title={VideoGPT: Video Generation using VQ-VAE and Transformers}, 
      author={Wilson Yan and Yunzhi Zhang and Pieter Abbeel and Aravind Srinivas},
      year={2021},
      eprint={2104.10157},
      archivePrefix={arXiv},
      primaryClass={cs.CV},
}

\newpage
\appendix

\section{Gyrokinetics}
\label{app:gyrokinetics}
This appendix provides a term-by-term decomposition of the gyrokinetic Vlasov equation introduced in \cref{eq:gyrovlasov}.
The presentation follows the standard $\delta f$ formulation~\citep{frieman1982nonlinear,Krommes_gyrokinetics_2012} in the local flux-tube limit, and the term numbering of \citet{PEETERS_GKW_2009}, then the quantities of interest are presented.

Each species satisfies the gyrokinetic equation:
\begin{equation}
\label{eq:gyrovlasov}
\underbrace{\displaystyle
\frac{\partial f}{\partial t} + (v_\parallel\,\Bb + \Bv_D)\cdot\nabla f
- \frac{\mu B}{m}\,\frac{\BB\cdot\nabla B}{B^{2}}\,\frac{\partial f}{\partial v_\parallel}}_{\text{Linear}}
\;+\;
\underbrace{\displaystyle
\smash{\Bv_E\cdot\nabla f}
\vphantom{\frac{\partial f}{\partial v_\parallel}}}_{\text{Nonlinear}}
\;=\; S,
\end{equation}
where $\Bb = \BB/B$ is the unit vector along the equilibrium magnetic field $\BB$, $\Bv_D$ the magnetic drift, $\Bv_E$ the $\BE\!\times\!\BB$ drift, and $S$ collects collisional, source terms and numerical dissipation. See Appendix \ref{app:gyrokinetics} for a detailed description of the gyrokinetic equation. The quantities of interest that can be extracted from gyrokinetics for downstream use are the electrostatic potentials, $\phi(t)$, and heat flux, $Q(t)$, both defined as integrals of $f$ (see Appendix \ref{app:gyrokinetics}), as well as their spectra $W(k_y)$ and $Q(k_y)$ respectively.

\subsection{The $\delta f$ decomposition}

The distribution function is split into a time-independent equilibrium Maxwellian and a fluctuating perturbation,
\begin{equation}
\label{eq:deltaf}
f \;=\; F_M + \delta f(v_\parallel, \mu, s, k_x, k_y;\, t).
\end{equation}
The equilibrium Maxwellian encodes the background density $n$ and temperature $T$ through
\begin{equation}
\label{eq:maxwellian}
F_M \;=\; \frac{n}{\left(2\pi T / m\right)^{3/2}}\;\exp\!\left(-\frac{m\,v^2}{2\,T}\right),
\end{equation}
where $m$ is the species mass.
The normalised radial gradients of $F_M$ yield the temperature and density gradient parameters $R/L_T$ and $R/L_n$ that enter the equilibrium drive and condition \ourmethod{} (\cref{eq:cond_drive}).

Substituting \cref{eq:deltaf} into the full gyrokinetic Vlasov equation and retaining terms to first order in $\delta f / F_M$ yields a closed evolution equation for $\delta f$.
The individual terms of the resulting right-hand side are detailed below.

\subsection{Right-hand side decomposition}

Substituting \cref{eq:deltaf} into \cref{eq:gyrovlasov} and separating linear from nonlinear contributions, the right-hand side groups into four categories: kinetic dynamics, energy drives, nonlinear advection, and numerical dissipation.
The $\Bv_E\!\cdot\!\nabla f$ term of \cref{eq:gyrovlasov} splits into the equilibrium drive~(V), which is linear in $\phi$, and the nonlinear advection~(III).
In the local flux-tube representation with spectral perpendicular coordinates $(k_x, k_y)$, the equation reads
\begin{equation}
\label{eq:rhs_decomposed}
\resizebox{0.95\textwidth}{!}{$\displaystyle
\begin{aligned}
\frac{\partial \delta f}{\partial t} \;=\;\; &
\underbrace{
\colorbox{termKinetic!15}{$\displaystyle
-v_\parallel\,\nabla_\parallel \delta f
\;-\; i(\Bk_\perp\!\cdot\!\Bv_D)\,\delta f
\;+\; \mu\,\nabla_\parallel B\;\frac{\partial \delta f}{\partial v_\parallel}
$}
}_{\text{Kinetic dynamics (I, II, IV)}}
\;\;\;
\underbrace{
\colorbox{termDrive!15}{$\displaystyle
-\Bv_E\!\cdot\!\nabla F_M
\;-\;\frac{Ze\,F_M}{T}\!\left(v_\parallel\,\nabla_\parallel\bar\phi + i(\Bk_\perp\!\cdot\!\Bv_D)\,\bar\phi\right)
$}
}_{\text{Energy drives (V, VII, VIII)}} \\[10pt]
&
\underbrace{
\colorbox{termNL!15}{$\displaystyle
\vphantom{\frac{Ze\,F_M}{T}}
-\Bv_E\!\cdot\!\nabla_\perp \delta f
$}
}_{\text{Nonlinear advection (III)}}
\;\;\;
\underbrace{
\colorbox{termDiss!15}{$\displaystyle
\vphantom{\frac{Ze\,F_M}{T}}
-\cD(\delta f)
$}
}_{\text{Dissipation}}
\end{aligned}
$}
\end{equation}
where $\bar\phi$ denotes the gyro-averaged electrostatic potential, $\Bv_D$ the magnetic drift velocity, $\Bv_E$ the $\BE\!\times\!\BB$ drift velocity, $Ze$ the species charge, and $\cD$ a numerical dissipation operator.
Each category is described below.

\vspace{0.5em}
\noindent\colorbox{termKinetic!20}{\textbf{Kinetic dynamics (I, II, IV).}}
These terms describe the collisionless motion of guiding centres in the equilibrium magnetic geometry.
Parallel advection~(I) represents streaming along the magnetic field.
The magnetic drift~(II) arises from curvature and $\nabla B$ drifts perpendicular to the field.
The mirror term~(IV) accounts for the parallel force exerted by gradients in the magnetic field magnitude, which leads to particle trapping in regions of low $B$. More explicitly, in $s\text{--}\alpha$ geometry,
\begin{equation}
\label{eq:cond_geom}
\nabla_\parallel
\;=\; \frac{1}{\,\colorbox{paramQ!20}{$\displaystyle \vphantom{\hat{s}} q$}\, R\,}\,\partial_{s},
\qquad
k_\perp^{2}
\;\propto\;
1 + \big(\colorbox{paramShat!20}{$\displaystyle \hat{s}$}\,s - \alpha\sin s\big)^{2}.
\end{equation}
so $q$ rescales parallel dynamics and $\hat{s}$ modulates the radial wavenumber along the field line.

\vspace{0.5em}
\noindent\colorbox{termDrive!20}{\textbf{Energy drives (V, VII, VIII).}}
The equilibrium drive~(V) couples the $\BE\!\times\!\BB$ drift to the radial gradient of the Maxwellian, providing the free-energy source for micro-instabilities.
It is through this term that the normalised gradients \paramRLTbox{} and \paramRLnbox{} enter the equation:
\begin{equation}
\label{eq:cond_drive}
\Bv_E\cdot\nabla F_{M}
\;=\; \Bv_E\cdot\hat{\mathbf{r}}\,F_{M}
\left(
\colorbox{paramRLn!20}{$\displaystyle \vphantom{\tfrac{R}{L_T}}\frac{R}{L_n}$}
\;+\;
\left(\frac{v^{2}}{v_{\mathrm{th}}^{2}} - \frac{3}{2}\right)
\colorbox{paramRLT!20}{$\displaystyle \frac{R}{L_T}$}
\right),
\end{equation}
where $v_{\mathrm{th}}$ is the thermal velocity.
The field drives~(VII, VIII) represent the linear response of the background distribution to the fluctuating electrostatic potential, coupling parallel streaming and magnetic drift to $\bar\phi$.

\vspace{0.5em}
\noindent\colorbox{termNL!20}{\textbf{Nonlinear advection (III).}}
The $\BE\!\times\!\BB$ advection of $\delta f$ by the fluctuating electric field is responsible for mode coupling, energy redistribution across spatial scales, and the development of saturated turbulence.
It is the computationally dominant term and the one that distinguishes nonlinear from linear gyrokinetics.

\vspace{0.5em}
\noindent\colorbox{termDiss!20}{\textbf{Dissipation.}}
A numerical dissipation operator $\cD$ is added for stability, acting along the parallel and velocity coordinates and as spectral hyper-viscosity in the perpendicular plane.
Its coefficients are chosen to damp grid-scale fluctuations without affecting the resolved physical scales.

\vspace{0.5em}
\noindent\textit{Remark.}
Term~VI in the convention of \citet{PEETERS_GKW_2009} is the neoclassical drive $-\Bv_D\!\cdot\!\nabla F_M$, which couples the magnetic drift to equilibrium gradients and is relevant only for rotating or neoclassical plasmas.
It is omitted in this work.

\subsection{Quasineutrality and the field solver}

The electrostatic potential $\phi$ is determined self-consistently from $\delta f$ through the gyrokinetic quasineutrality condition.
In Fourier space, for each perpendicular wavevector $\Bk_\perp = (k_x, k_y)$ and parallel position $s$,
\begin{equation}
\label{eq:quasineutrality}
\sum_a \frac{Z_a^2\, n_a}{T_a}\left(\Gamma_0^a - 1\right)\hat\phi(\Bk_\perp, s) \;=\; \sum_a Z_a \int J_0^a\;\delta f_a\;B\;\mathrm{d}v_\parallel\;\mathrm{d}\mu,
\end{equation}
where the sum runs over species $a$ with charge number $Z_a$, density $n_a$, and temperature $T_a$.
The operator $J_0^a = J_0(k_\perp\rho_a)$ is the zeroth-order Bessel function that performs the gyro-average, corresponding to the operator $\mathbf{J_0}$ appearing in \cref{eq:integrals}.
The quantity $\Gamma_0^a = I_0(b_a)\,e^{-b_a}$ with $b_a = k_\perp^2\rho_a^2$ is the velocity-space-integrated gyro-average, where $I_0$ is the modified Bessel function of the first kind and $\rho_a$ is the species gyroradius.

\Cref{eq:quasineutrality} is algebraic in $\hat\phi$ and is solved at each evaluation of the right-hand side.
In the adiabatic electron approximation used for the dataset in this work, only the ion species is evolved kinetically and the electron response is replaced by a Boltzmann relation, $\delta n_e / n_e = e\,\phi / T_e$.
This simplifies the left-hand side of \cref{eq:quasineutrality} and removes the need to resolve the fast electron time scales, while retaining the essential ion-temperature-gradient-driven turbulence.

\subsection{Integrals and diagnostics}
\label{sec:integrals}
The quantities of interest are the electrostatic potential $\bm{\phi}(x, s, y)$ and the scalar heat flux $Q\in\mathbb{R}$. Following \citet{PEETERS_GKW_2009}, both are velocity-space integrals of $f$,
\begin{equation}
\label{eq:integrals}
\bm{\phi} \;=\; \mathbf{A}\int \mathbf{J_{0}}\,\bm{f}\;\mathrm{d}v_\parallel\mathrm{d}\mu,
\qquad
Q \;=\; \int \mathbf{C}\int v^{2}\,\bm{\phi}\,\bm{f}\;\mathrm{d}v_\parallel\mathrm{d}\mu\;\mathrm{d}x\,\mathrm{d}y\,\mathrm{d}s,
\end{equation}
where $\mathbf{A},\mathbf{C}\in\mathbb{R}^{x\times s\times y}$ collect geometric and operating coefficients, $v^{2}$ is the pointwise kinetic energy, and $\mathbf{J_{0}}$ is the zeroth-order Bessel envelope.
Turbulence is diagnosed through binormal-direction wavespace spectra
\begin{equation}
\label{eq:ky_spectra}
W(k_y) \;=\; \sum_{s,\, k_x} \big|\hat{\bm{\phi}}(k_x, s, k_y)\big|^{2},
\qquad
Q(k_y) \;=\; \sum_{v_\parallel,\mu,\, s,\, k_x} \bm{Q}(v_\parallel, \mu, s, k_x, k_y),
\end{equation}
with $\hat{\bm{\phi}}$ the Fourier-space potential and $\bm{Q}$ the heat-flux field prior to the outer integral of \cref{eq:integrals}. In the saturated regime the amplitude of radial transport is regulated by zonal flows at $k_y = 0$ \citep{Itoh_ZF_06}. The quantities of practical interest are time-averaged: $\bar{Q}$, $Q(k_y)$, and $W(k_y)$.

\section{Data generation}
\label{app:data}

\begin{figure}[h]
    \centering
    \includegraphics[width=\linewidth]{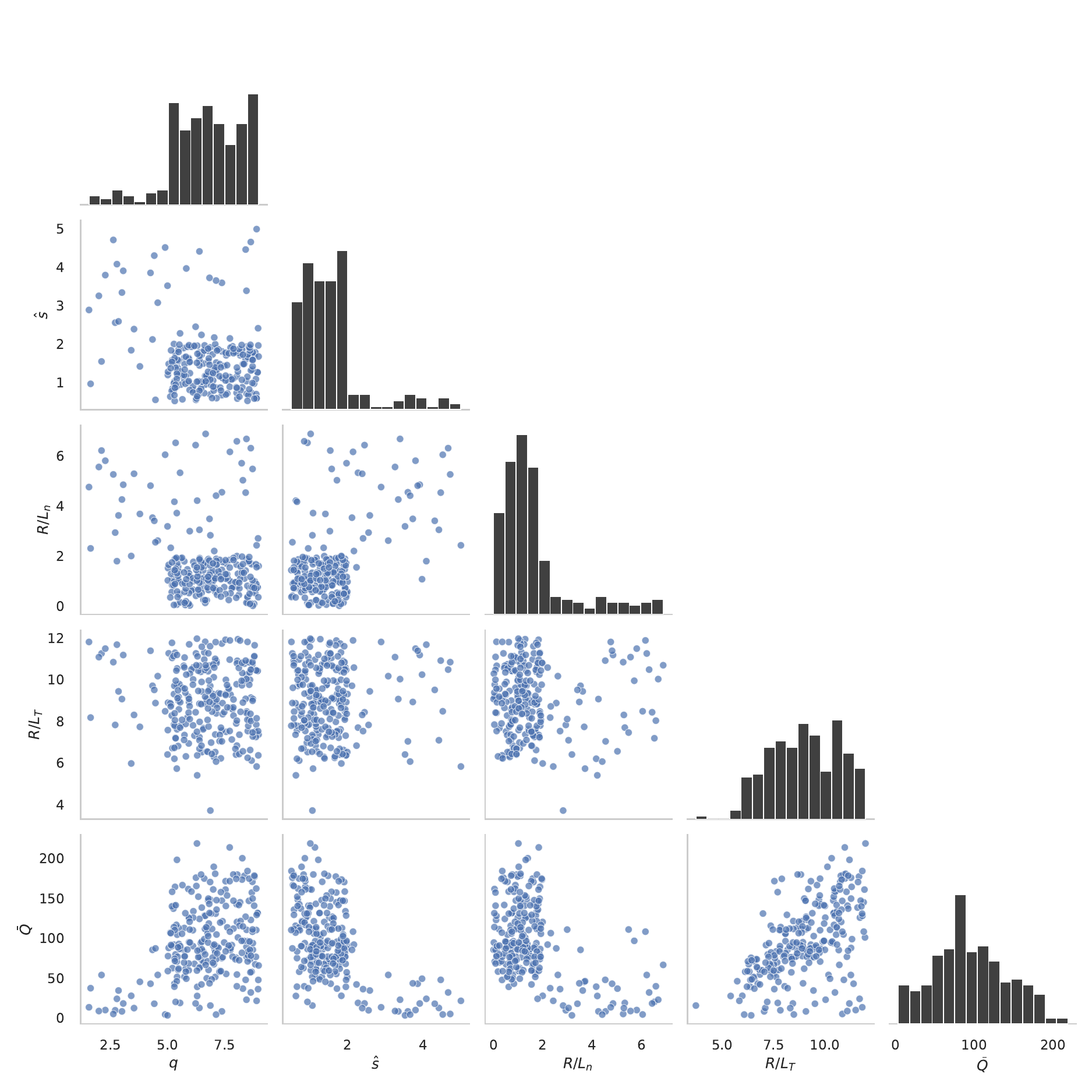}
    \caption{Distribution of the four operating parameters $\hat{s}$, $q$, $R/L_n$, $R/L_T$ and the resulting mean heat flux $\bar{Q}$ across the dataset, after the two-pass sweep. Reproduced from \citet{paischer2025gyroswin}.}
    \label{fig:param_dist}
\end{figure}

We reuse the dataset introduced in \citet{paischer2025gyroswin}. We summarise it here.
All simulations are cyclone-base-case ion-temperature-gradient runs \citep{dimits_cbc_2000} generated with GKW \citep{PEETERS_GKW_2009}, parameterised by four dimensionless numbers from \cref{sec:params}.
Data collection is performed in two passes.
The first pass samples $R/L_T \in [3, 12]$, $R/L_n \in [1, 7]$, $q \in [1, 9]$, $\hat{s} \in [0.5, 5]$, together with an initial-condition noise amplitude in $[10^{-5}, 10^{-3}]$ and initial shape in $\{\sin, \cos, \text{random}\}$. Out of 100 runs, only about half reach saturation. Following the sparsity of turbulent points in \cref{fig:param_dist}, the second pass narrows the ranges to $R/L_T \in [6, 12]$, $R/L_n \in [0, 2]$, $q \in [5, 9]$, $\hat{s} \in [0.5, 2]$, reducing the stabilising factors; all 200 runs of the second pass develop turbulence.
After filtering, the dataset consists of roughly 250 saturated nonlinear trajectories. We hold out three for validation and six as an in-distribution test split, and retain 241 trajectories for training. Linear counterparts of each configuration are generated in parallel for the reduced-order baselines.

\section{Single-parameter sensitivity scans}
\label{app:scan}

This appendix specifies the protocol for the $\bar Q$ scans of \cref{fig:param_scans}.

\paragraph{Operating points.}
Starting from a fixed strongly turbulent baseline ($R/L_T{=}10.17$, $R/L_n{=}2.61$, $\hat s{=}3.08$, $q{=}4.57$), we sweep one of $\{R/L_T,\,R/L_n,\,\hat s,\,q\}$ at a time on a $K{=}6$-point linspace covering a wide range of that parameter (even outside the training regime), holding the other three fixed. The 24 scan points are disjoint from the training and test trajectories of \cref{app:data}, so each scan is a single-axis OOD probe. \cref{tab:scan_points} lists every operating point, the swept value, and the GT saturated heat flux.

\paragraph{Reference solver.}
GT $\bar Q$ at each scan point is obtained from gyaradax \citep{galletti2026gyaradax} (a GKW port) in the adiabatic-electron limit. Each run integrates for $318$ time units from a $\cos^2$-shaped IC. We report the saturated mean as the average flux over the last $96$ time units. Three interior scan points hit a numerical NaN under the exact baseline parameters (sensitive boundary-of-aliasing configurations); we re-ran those with $\Delta\!\sim\!1\!-\!5\%$ jitter on the swept value (kept inside the linspace cell) and report the first stable run. One scan point ($\hat s{=}5.00$) failed to converge under any jitter inside the cell and is reported as NaN.

\paragraph{Plot.}
\cref{fig:param_scans} reports the mean of the $64$ model samples (green $\diamond$ markers) and the per-sample distribution as a faint cloud at the same $x$. GT is the saturated mean from gyaradax (dark $\circ$ markers). With a gray shadow we report the histogram of training samples on flux and parameters.

\begin{table}[h]
\centering
\small
\caption{All 24 scan points and their gyaradax saturated heat flux $\bar Q_{\mathrm{GT}}$. Each block sweeps one operating parameter across the empirical min/max of the training corpus while the other three are held at the baseline.}
\label{tab:scan_points}
\setlength{\tabcolsep}{5pt}
\renewcommand{\arraystretch}{1.05}
\begin{tabular}{c r r r r r r r r}
\toprule
\multirow{2}{*}{point} & \multicolumn{2}{c}{$R/L_T$} & \multicolumn{2}{c}{$R/L_n$} & \multicolumn{2}{c}{$\hat s$} & \multicolumn{2}{c}{$q$} \\
\cmidrule(lr){2-3} \cmidrule(lr){4-5} \cmidrule(lr){6-7} \cmidrule(lr){8-9}
& swept value & $\bar Q_{\mathrm{GT}}$ & swept value & $\bar Q_{\mathrm{GT}}$ & swept value & $\bar Q_{\mathrm{GT}}$ & swept value & $\bar Q_{\mathrm{GT}}$ \\
\midrule
0 & $1.060$  & $0.00$  & $0.004$ & $45.69$           & $0.510$ & $101.33$          & $1.000$ & $0.00$ \\
1 & $3.244$  & $0.00$  & $1.401$ & $51.71$           & $1.408$ & $83.35$           & $2.598$ & $25.35$ \\
2 & $5.428$  & $3.12$  & $2.798$ & $50.52$           & $2.306$ & $78.97$           & $4.196$ & $43.75$ \\
3 & $7.612$  & $18.38$ & $4.196$ & $41.88$           & $3.204$ & $47.05$           & $5.794$ & $58.23$ \\
4 & $9.796$  & $43.98$ & $5.593$ & $29.55$           & $4.102$ & $33.57$           & $7.392$ & $50.33$ \\
5 & $11.980$ & $85.81$ & $6.990$ & $18.56$           & $5.000$ & NaN               & $8.990$ & $48.41$ \\
\bottomrule
\end{tabular}
\end{table}


\section{Training and architecture details}
\label{app:training}

This appendix collects the architectural and optimisation
hyperparameters of \ourmethod{} and the three generative baselines
introduced in \cref{sec:experiments}. Unless stated otherwise, all
models share the same Swin5D backbone of \cref{sec:method} and differ
only in their bottleneck and the way the operating parameters $\Bc$
enter. In \cref{tab:ae_recon_combined} the reconstruction RMSE of the autoencoders is listed, as well as the count of the parameters.

\subsection{Swin5D autoencoder}
\label{app:training_ae}

The encoder $E_\psi$ and decoder $D_\psi$ use a hierarchical Swin5D backbone of depth 8 with 16 attention heads per layer. Attention is computed locally over non-overlapping windows of size $M_{v_\parallel}\!\times\!M_s\!\times\!M_x\!\times\!M_y = 4\!\times\!4\!\times\!9\!\times\!4$, alternating standard and shifted-window partitions across consecutive blocks. The $\mu$ axis is folded into the channel dimension after a learned 1D positional embedding rather than attended over directly, which keeps attention 4-dimensional while still resolving the variable-amplitude $\mu$ shells. The bottleneck consists of two Transformer layers with four attention heads operating at the coarsest resolution after a linear down-projection to channel width $d_z = 256$. Both $E_\psi$ and $D_\psi$ are unconditional. Training uses an MSE reconstruction loss on the 2-channel real-space representation of $f$ described in \cref{sec:method}.

\subsection{GyroSwin baselines: cold vs.\ warm$^{\ast}$}
\label{app:gyroswin_cold_vs_warm}

\Cref{tab:downstream} lists two GyroSwin variants. $\text{GyroSwin (warm)}^{\ast}$ is the public checkpoint of \citet{paischer2025gyroswin} \footnote{\url{https://huggingface.co/datasets/ml-jku/gyroswin_cbc_id_ood}}. GyroSwin (cold) is a retraining of the same hierarchical Swin5D architecture by us, with the differences summarised in \cref{tab:gyroswin_cold_vs_warm}.

\textbfp{Two-stage training}
Both checkpoints follow the same two-stage protocol: a pretraining stage that supervises one-step prediction of $f$ and $\phi$, and a posttraining stage that warm-starts from it, freezes everything outside the conditional flux head (\texttt{params\_to\_include=[flux]}), and supervises only the time-averaged scalar heat flux $\bar{Q}$ via a loss schedule that ramps $w_{f}, w_{\phi}\rightarrow 0$ and $w_{\bar{Q}}\rightarrow 1$.

\textbfp{Cold vs.\ warm trajectory window}
The warm$^{\ast}$ model is pretrained on windows starting after the linear-to-nonlinear transient, and is rolled out from a saturated snapshot. GyroSwin (cold) is pretrained on the full trajectories, exposing it to the entire ramp-up, and at inference is rolled out autoregressively from a small random perturbation through the transient until saturation, skipping the need for using a numerical solver for starting the rollout.

\textbfp{Architecture alignment with the autoencoder}
We bring the GyroSwin backbone in line with the Swin5D autoencoder of \cref{sec:method}, so the surrogate trained from the cold start uses the same stability axes as the AE that drives \ourmethod{}. Concretely, against the warm$^{\ast}$ backbone we change \texttt{LayerNorm}~$\rightarrow$ \texttt{RMSNorm}, \texttt{film}~$\rightarrow$ adaLN-Zero (\texttt{dit}), enable \texttt{gated\_attention} and \texttt{qk\_norm}, and increase the head dimension from $16$ to $32$ (\texttt{num\_heads} $64{\rightarrow}32$; $d_{z}{=}1024$ unchanged). The 5D patch and window partition and the network depth are unchanged.

\textbfp{Normalization}
The warm$^{\ast}$ model uses a single global $z$-score across all fields. GyroSwin (cold) uses the per-field, $\mu$-decoupled scheme of \cref{sec:method} (\texttt{norm\_decouple\_mu=true}, with $f$ aggregated over $(\mu, s, v_\parallel, x, y)$ and $\phi$ over $(s, x, y)$).

\textbfp{Optimization}
Pretraining runs 500 epochs on $32{\times}4$ H100 GPUs at effective batch size $1024$, AdamW with $\eta{=}2.1{\times}10^{-4}$, cosine schedule to $\eta_{\min}{=}10^{-6}$, gradient clip $0.1$, bf16. Posttraining runs $200$ further epochs (501-700) with $\eta{=}5{\times}10^{-5}$, clip $0.1$, and the backbone frozen.

\begin{table}[t]
\centering\small
\caption{GyroSwin (cold) vs.\ $\text{GyroSwin (warm)}^{\ast}$. Fields
identical between the two (patch/window, depth, $d_z{=}1024$, two-stage
protocol, fluxavg loss schedule, posttrain \texttt{flux\_conditioning})
are omitted.}
\label{tab:gyroswin_cold_vs_warm}
\begin{tabular}{lll}
\toprule
& GyroSwin (cold) & GyroSwin (warm)$^{\ast}$ \\
\midrule
trajectory window (pretrain) & full ($\texttt{offset}{=}0$) & saturated ($\texttt{offset}{=}80$) \\
inference IC                 & random perturbation          & saturated snapshot \\
\midrule
norm                         & RMSNorm                      & LayerNorm \\
modulation                   & adaLN-Zero (DiT)             & FiLM \\
gated attention / QK-norm    & yes / yes                    & no / no \\
\#heads / head-dim           & 32 / 32                      & 64 / 16 \\
field normalization          & per-field, $\mu$-decoupled   & global $z$-score \\
\midrule
nodes / batch & 32 / 1024 / (GH200$\times 4$)                 & 4 / 128 / (H100$\times 4$) \\
pretrain LR / clip           & $2.1{\times}10^{-4}$ / $0.1$ & $3{\times}10^{-4}$ / $0.5$ \\
posttrain LR / clip          & $5{\times}10^{-5}$ / $0.1$   & $3{\times}10^{-4}$ / $0.5$ \\
pretrain / posttrain epochs  & 500 / 200                    & 500 / 200 \\
\bottomrule
\end{tabular}
\end{table}

\subsection{Latent generative models}
\label{app:training_gen}

All four generative models reuse the autoencoder of \cref{app:training_ae}; they differ in their bottleneck and in how $\Bc$ is injected.

\textbfp{VAE} A reparameterised Gaussian bottleneck with channel width $d_z = 256$. The KL weight is annealed cyclically from $0$ to $0.2$ over four cycles to mitigate posterior collapse, and the posterior log-variance is clamped to $[-20, 20]$ to avoid degenerate encodings. The encoder is unconditional; the decoder receives $\Bc$ through per-block adaptive layer normalization (adaLN)~\citep{peebles_dit_2023}. At inference, $z\sim\mathcal{N}(0, I)$ is decoded with the test-time $\Bc$.

\textbfp{VQ-VAE} A vector-quantised bottleneck with an EMA-updated codebook of $8192$ entries, embedding dimension $256$, and commitment loss weight $0.25$. The conditioning structure mirrors the VAE (unconditional encoder, adaLN-conditioned decoder). At inference, code indices are drawn independently from the empirical codebook distribution estimated on the training set, and decoded jointly with $\Bc$. For the vector quantization we were using the implementation from \texttt{vector-quantize-pytorch}\footnote{\url{https://github.com/lucidrains/vector-quantize-pytorch}}.

\textbfp{VQ-VAE + Transformer} The same VQ-VAE bottleneck, paired with a $12$-layer, $16$-head GPT-style causal Transformer with $d_\text{model} = 1024$ trained on the code sequences. The cross-entropy objective uses label smoothing $0.1$, and $\Bc$ is injected per layer via FiLM~\citep{perez2018film}. Sampling is autoregressive with KV-caching, and the resulting index sequence is decoded by the frozen VQ-VAE decoder under the same $\Bc$.

\textbfp{\ourmethod{}} A plain autoencoder (no KL, no quantisation) with an unconditional encoder and decoder, and a DiT that operates directly on the bottleneck tokens with no additional patching. The flow time $t$ and the four entries of $\Bc$ are each lifted through a sinusoidal embedding and a shared MLP to a joint conditioning vector that modulates each DiT block via adaLN. Residual gates are zero-initialised so the DiT begins training as the identity. Architectural hyperparameters of the DiT are: depth $12$, $16$ attention heads, $d_\text{model} = 1024$, MLP ratio $2.0$, token input/output dimension $d_z = 256$ (matching the AE bottleneck), and drop-path $0.1$. The flow time $t$ and the four entries of $\Bc$ are each lifted to $128$ dimensions through a sinusoidal embedding ($\omega_k = 10^{-4k/K}$) followed by a SiLU MLP, concatenated, and broadcast to every block as the adaLN modulation signal.


\begin{table}[t]
\centering
\caption{Reconstruction RMSE in physical (denormalised) units, evaluated for each autoencoder backbone underlying the latent generative models. Parameter counts in millions (M). ID and OOD entries denote the mean $\pm$ standard deviation across trajectories of the corresponding test split.}
\label{tab:ae_recon_combined}
\small
\begin{tabular}{lccccc}
\toprule
\multirow{2}{*}{\textbf{Method}} & \multirow{2}{*}{\textbf{Params (M)}} & \multicolumn{2}{c}{${Recon}_{\mathrm{RMSE}}\downarrow$} & \multicolumn{2}{c}{$\bar{Q}_{\mathrm{RMSE}}\downarrow$} \\
\cmidrule(lr){3-4} \cmidrule(lr){5-6}
 &  & \textbf{ID} & \textbf{OOD} & \textbf{ID} & \textbf{OOD} \\
\midrule
AE & 217.6 & $0.753_{\pm 0.175}$ & $0.786_{\pm 0.257}$ & $4.76_{\pm 2.15}$ & $7.12_{\pm 7.35}$ \\
VAE & 217.9 & $0.839_{\pm 0.171}$ & $0.869_{\pm 0.249}$ & $3.82_{\pm 0.654}$ & $6.92_{\pm 6.39}$ \\
VQ-VAE & 225.8 & $0.888_{\pm 0.187}$ & $0.924_{\pm 0.273}$ & $21.6_{\pm 9.7}$ & $23.5_{\pm 13.4}$ \\
\bottomrule
\end{tabular}
\end{table}

\subsection{Optimisation}
\label{app:training_optim}

The DiT and the autoregressive prior are trained with AdamW; the Swin5D autoencoders (AE, VAE, VQ-VAE) use Adam (PyTorch \texttt{weight\_decay} added to the gradient, not decoupled). All runs use $\beta_1 = 0.9$, $\beta_2 = 0.999$, $\epsilon = 10^{-8}$, weight decay $10^{-6}$, gradient clipping at norm $1.0$, and bf16 mixed precision. Learning rates and schedules differ across components and reflect the regimes in which each was found stable (\cref{tab:training_hparams}). \Cref{tab:training_hparams} also lists batch sizes and total epochs.

\begin{table}[h]
\centering
\caption{Optimisation hyperparameters for the autoencoder, the autoregressive prior over VQ-VAE codes, and the DiT used in \ourmethod{}. All runs use AdamW, weight decay $10^{-6}$, gradient clipping at $1.0$, and bf16 mixed precision.}
\label{tab:training_hparams}
\small
\begin{tabular}{lcccc}
\toprule
\textbf{Component} & \textbf{Batch size} & \textbf{Epochs} &
\textbf{Peak lr} & \textbf{Schedule} \\
\midrule
Swin5D autoencoder (all variants) & 512 & 400  & $3\!\cdot\!10^{-4}$ & cosine \\
VQ-VAE Transformer prior          & 128 & 1000 & $1\!\cdot\!10^{-3}$ & OneCycle \\
\ourmethod{} DiT                  & 128 & 1000 & $5\!\cdot\!10^{-4}$ & cosine \\
\bottomrule
\end{tabular}
\end{table}

For all autoencoders we observe early-training instabilities driven by large $\mu$-shell amplitude imbalance; the per-shell normalization of \cref{sec:method} eliminates them. Without it, training of the larger Swin5D variants diverges within the first epoch. QK normalization and gated attention contribute smaller but consistent improvements at scale, in line with the LLM literature~\citep{henry2020querykey,qiu2025gated}.

\subsection{Flow-matching specifics}
\label{app:training_fm}

The DiT in \ourmethod{} is trained with the rectified flow-matching objective of \cref{eq:fm_loss}. As detailed in \cref{app:generative}, we draw the integration time from a logit-normal distribution ($\tau\sim\mathcal{N}(0, 1)$, $t = \sigma(\tau)$) and pair noise with data within each minibatch via the optimal-transport assignment of \cref{eq:ot_coupling}, solved with the Hungarian algorithm at $\mathcal{O}(B^3)$ cost. Latents are rescaled by their average per-element standard deviation $\sigma_z$, estimated once over the training set, before flow matching. At inference we use $N = 15$ explicit-Euler steps on a uniform grid in $[0, 1]$.

\subsection{Compute}
\label{app:training_compute}

All models are trained on NVIDIA GH200 GPUs (the \texttt{gracehopper} cluster, 4 GPUs per node), using bf16 mixed precision throughout. \Cref{tab:training_compute} summarises the per-component compute budget. Inference cost per sample is reported in \cref{tab:downstream}.

\begin{table}[h]
\centering
\caption{Compute used for each training run, on NVIDIA GH200 120GB (\texttt{gracehopper} cluster, 4 GPUs/node, bf16 mixed precision throughout). Wall-clock is end-to-end. The DiT and AR runs are single-GPU; the autoencoders use multi-node data parallelism (DeepSpeed ZeRO-2 for the AE, PyTorch DDP for the VAE/VQ-VAE).}
\label{tab:training_compute}
\small
\begin{tabular}{lccccc}
\toprule
\textbf{Run} & \textbf{GPUs} & \textbf{Parallelism}
& \textbf{Wall-clock} & \textbf{GPU-hours} & \textbf{Epochs} \\
\midrule
\ourmethod{} AE                & 32 & DeepSpeed ZeRO-2 & 19 h  & $\sim\!600$ & 400 \\
VAE                            & 64 & DDP             & 11 h  & $\sim\!695$ & 400 \\
VQ-VAE                         & 64 & DDP             &  9 h  & $\sim\!580$ & 400 \\
VQ-VAE Transformer prior (AR)  & 1  & --              & 19 h  & $\sim\!19$  & 1000 \\
\ourmethod{} DiT               & 1  & --              & 24 h  & $\sim\!24$  & 1000 \\
\bottomrule
\end{tabular}
\end{table}

\section{Generative model details}
\label{app:generative}

This appendix expands the flow-matching construction of \cref{sec:method} and gives the pragmatic variance-reduction choices that make training on 5D plasma latents stable.

\subsection{Stochastic interpolants and rectified flow}
The stochastic-interpolant framework of \citet{albergo2023stochastic} bridges two arbitrary densities $\rho_0$ and $\rho_1$ on $\mathbb{R}^d$ exactly in finite time through a time-indexed interpolant
\begin{equation}
\label{eq:sit_general}
x_t \;=\; \alpha(t)\,x_0 \;+\; \beta(t)\,x_1 \;+\; \gamma(t)\,z, \qquad t\in[0, 1],
\end{equation}
with $x_0\sim\rho_0$, $x_1\sim\rho_1$, and $z\sim\mathcal{N}(0, I)$ an independent latent, subject to $\alpha(0)=\beta(1)=1$ and $\alpha(1)=\beta(0)=\gamma(0)=\gamma(1)=0$. The time-dependent density of $x_t$ satisfies a first-order transport equation together with a family of forward and backward Fokker--Planck equations with tunable diffusion, so the same marginal law can be realised either as an ODE or as an SDE of matching diffusion coefficient. The drift coefficients entering these equations are the unique minimisers of simple quadratic objectives estimable from samples of $\rho_0$ and $\rho_1$.

We instantiate the \emph{spatially linear one-sided} interpolant of \citet[\S 4.4]{albergo2023stochastic}, which takes $\rho_0 = \mathcal{N}(0, I)$, $\alpha(t)=1-t$, $\beta(t)=t$, $\gamma\equiv 0$, and selects the ODE branch with zero diffusion. \Cref{eq:sit_general} then collapses to the straight-line path of \cref{eq:fm_path}, and the target drift reduces to the rectified-flow velocity $u^\star(x_t\mid x_0, x_1) = x_1 - x_0$ of \citet{liu2022}. In the flow-matching formulation of \citet{lipmanflow}, this is the probability path induced by a Gaussian prior and an independent-pair coupling; we extend it below with an OT coupling that tightens this pairing.

\subsection{Logit-normal time schedule}
Uniform sampling of $t\sim\mathcal{U}(0, 1)$ oversamples the endpoints, where the target velocity is already close to $x_1 - x_0$ (near $t{=}1$) or close to the prior mean (near $t{=}0$). The intermediate regime, where the network has to disambiguate mode structure from noise, is in contrast undersampled. Following \citet{esser2024scalingrectifiedflowtransformers}, we draw $\tau\sim\mathcal{N}(0, 1)$ and set $t = \sigma(\tau) = (1+\mathrm{e}^{-\tau})^{-1}$, which concentrates training density in $t\approx 0.5$ and leaves the endpoints with lighter sampling. In practice we observe a clear reduction in training-loss variance and faster convergence at the same compute.

\subsection{Optimal-transport minibatch coupling}
The independent pairing $(z_0^{(i)}, z_1^{(i)})$ sampled within a minibatch induces paths that frequently cross and produce high-variance velocity targets. Following \citet{tong2024improvinggeneralizingflowbasedgenerative}, we permute prior samples against data latents by solving the assignment problem
\begin{equation}
\label{eq:ot_coupling}
\pi^{\star} \;=\; \arg\min_{\pi\in\mathfrak{S}_{B}}\; \sum_{i=1}^{B} \big\|\, z_{0}^{(\pi(i))} - z_{1}^{(i)} \,\big\|_{2},
\end{equation}
where $\mathfrak{S}_{B}$ is the symmetric group on $B$ elements, and use the matched pairs $(z_0^{(\pi^\star(i))}, z_1^{(i)})$ in place of the independent ones. In practice we flatten each latent to a vector, build the Euclidean cost matrix, and solve \cref{eq:ot_coupling} with the Hungarian algorithm at $\mathcal{O}(B^3)$ cost, which is negligible compared to a forward pass of $v_\theta$ at small batch sizes. This pairing approximates the optimal-transport coupling between the empirical prior and empirical data measures, straightens the resulting paths, and shortens inference-time trajectories at fixed $N$.

\subsection{Inference}
At inference time we draw $z \sim \mathcal{N}(0, I)$ and integrate $\dot z = v_{\theta}(z, t, \Bc)$ with an explicit Euler scheme,
\begin{equation}
\label{eq:fm_inference}
z_{k+1} = z_{k} + \Delta t\, v_{\theta}(z_{k}, t_{k}, \Bc), \qquad t_{k} = k/N,\; \Delta t = 1/N,
\end{equation}
on a uniform grid of $N$ steps in $[0, 1]$.
The combination of rectified-flow straight paths, logit-normal time sampling, and OT-coupled training is known to shorten inference trajectories at fixed sample quality \citep{liu2022, esser2024scalingrectifiedflowtransformers, tong2024improvinggeneralizingflowbasedgenerative}.

\subsection{Choice of integration steps}
\label{app:euler_steps}
\begin{figure}[t]
\centering
\includegraphics[width=0.7\textwidth]{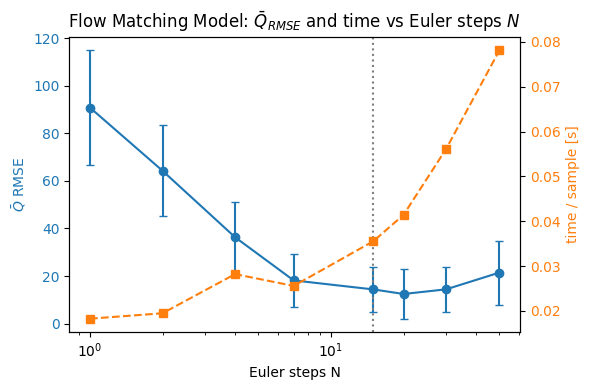}
\caption{$\bar{Q}_\text{RMSE}$ (blue, left axis) and per-sample wall-clock time (orange, right axis) versus the number of explicit-Euler integration steps $N$. Error bars denote standard deviation over the validation split. The dotted line marks our operating point at $N=15$.}
\label{fig:euler_sweep}
\end{figure}

\Cref{fig:euler_sweep} reports $\bar{Q}_\text{RMSE}$ and per-sample wall-clock time as a function of the number of explicit-Euler steps $N$ used to integrate $\dot z = v_\theta(z, t, \Bc)$ at inference. Accuracy improves rapidly up to $N{\approx}7$, reaches its minimum around $N{=}15$, and degrades slightly with larger $N$ as the spread across samples grows, while runtime increases roughly linearly. We adopt $N=15$ throughout \cref{sec:experiments}: it sits at the elbow of the accuracy curve and keeps generation below $35$\,ms per sample.

\section{Warm starts and the Kolmogorov-Smirnov statistic}
\label{app:ks}
\label{app:warmstart}

This appendix specifies the construction of the warm-start divergence reported in \cref{tab:warmstart_fid}\subref{tab:warmstart_ad_panel} and used as the $y$-axis of \cref{fig:fid_ad_corr}.

\subsection{Setup}
Fix a held-out operating point $\Bc$ and let $S(t) \in \mathbb{R}^{n_{k_y}}$ denote one of the binormal spectra of \cref{sec:gyrokinetics}, either $W(k_y, t)$ or $Q(k_y, t)$. Saturated turbulence is statistically stationary, so along a sufficiently long trajectory the empirical distribution of $\{S(t)\}$ converges to a $\Bc$-dependent stationary law $\mathbb{P}_{\Bc}$ on $\mathbb{R}^{n_{k_y}}$. We compare two empirical estimates of $\mathbb{P}_{\Bc}$:
\begin{itemize}[leftmargin=1.5em,itemsep=2pt,topsep=2pt]
  \item the \emph{warm-side} sample $\mathcal{S}^{w}_{\Bc}$ obtained by initializing gyaradax from generated fields $\hat f^{(k)} \sim p_{\theta}(\,\cdot\mid \Bc)$ drawn from \ourmethod{} and integrating for $T$ steps;
  \item the \emph{reference} sample $\mathcal{S}^{r}_{\Bc}$ obtained from the post-saturation tail of a long trajectory at the same $\Bc$.
\end{itemize}
Under the null hypothesis that $\hat f^{(k)}$ already lies on the attractor of $\mathbb{P}_{\Bc}$, the two samples are i.i.d.\ from the same law and any statistical test should fail to reject equality.

\subsection{Two-sample Kolmogorov-Smirnov statistic}
We score the agreement between $\mathcal{S}^{w}_{\Bc}$ and $\mathcal{S}^{r}_{\Bc}$ with the two-sample Kolmogorov-Smirnov (KS) statistic \citep{Massey1951}, the classical non-parametric measure of distributional discrepancy. Given samples $\mathcal{S}^{w} = \{x_{i}\}_{i=1}^{n}$ and $\mathcal{S}^{r} = \{y_{j}\}_{j=1}^{m}$ with empirical CDFs $F_n^{w}(x) = \tfrac{1}{n}\sum_i \mathbf{1}_{x_i \le x}$ and $F_m^{r}(x) = \tfrac{1}{m}\sum_j \mathbf{1}_{y_j \le x}$, the statistic is the supremum of their pointwise gap,
\begin{equation}
\label{eq:ks_definition}
D_{n,m}\!\big(\mathcal{S}^{w}, \mathcal{S}^{r}\big)
\;=\; \sup_{x \in \mathbb{R}}\, \big| F_n^{w}(x) \,-\, F_m^{r}(x) \big| \;\in\; [0, 1].
\end{equation}
$D_{n,m} = 0$ corresponds to identical empirical CDFs and $D_{n,m} = 1$ to disjoint supports. Under the null $\mathbb{P} = \mathbb{Q}$, the rescaled statistic $\sqrt{nm/(n{+}m)}\,D_{n,m}$ converges weakly to the Kolmogorov distribution and admits a distribution-free $p$-value; we report $D$ rather than the $p$-value because $p$ collapses below numerical resolution at our pooled sample sizes ($\sim 10^{4}$). $D$ requires no kernel or bandwidth choice, is monotone in distributional discrepancy, and is computed in $\mathcal{O}((n+m)\log(n+m))$ via sorted-merge of the two samples. We use \texttt{scipy.stats.ks\_2samp} for the per-mode evaluation.

\subsection{Restart pooling}
A single warm trajectory yields a strongly auto-correlated time-series whose empirical mode-marginal sits at one realisation of the attractor and may differ from $\mathbb{P}_{\Bc}$ by an amount comparable to the inter-trajectory variation we wish to measure. To reduce this within-condition variance we draw $R = 5$ independent generations $\{\hat f^{(r)}\}_{r=1}^{R}$ from \ourmethod{} at the same $\Bc$, integrate each for $T$ steps with gyaradax, and pool the post-saturation halves along the time axis,
\begin{equation}
\label{eq:warm_pool}
\mathcal{S}^{w}_{\Bc}
\;=\; \bigcup_{r=1}^{R}\, \big\{\, S^{(r)}(t) \,:\, t \in [\,T/2,\,T\,]\,\big\},
\qquad
\big|\mathcal{S}^{w}_{\Bc}\big| \;=\; R\, T/2.
\end{equation}
Pooling along time is justified by the stationarity of $\mathbb{P}_{\Bc}$. We additionally compute $D$ once per restart against the same reference, and report the cross-restart standard deviation as a within-condition uncertainty for the scatter of \cref{fig:fid_ad_corr}.

\subsection{Per-mode aggregation}
For vector spectra we treat $S(t) \in \mathbb{R}^{n_{k_y}}$ component-wise. At binormal mode index $\ell \in \{1, \ldots, n_{k_y}\}$ the scalar warm-side and reference samples are $\{S^{w}_{\ell}(t)\}$ and $\{S^{r}_{\ell}(t)\}$ with empirical CDFs $\hat F^{w}_{\ell}, \hat F^{r}_{\ell}$, and the per-mode KS statistic is
\begin{equation}
\label{eq:ks_permode}
D_\ell \;=\; \sup_{x \in \mathbb{R}}\,\big|\hat F^{w}_{\ell}(x) - \hat F^{r}_{\ell}(x)\big| \in [0, 1],
\end{equation}
the form of \cref{eq:ks_definition} applied to a single mode. The headline reported in \cref{tab:warmstart_fid}\subref{tab:warmstart_ad_panel} is the arithmetic mean over modes,
\begin{equation}
\label{eq:ks_modemean}
\overline{D}_{S}(\Bc)
\;=\; \frac{1}{n_{k_y}}\sum_{\ell=1}^{n_{k_y}} D_\ell.
\end{equation}
Per-mode aggregation makes the statistic insensitive to the absolute amplitude scale across $k_y$, which would otherwise be dominated by the high-energy low-$k_y$ end. Each $D_\ell$ is computed via \texttt{scipy.stats.ks\_2samp}.

\section{Distributional metrics: Wasserstein and \ourfid{}}
\label{app:fid}

This appendix lists the four GyroSwin U-Net depths we tracked during development, two of which are reported in the main paper, and addresses the statistical reliability of \ourfid{} at the sample counts we use.

\subsection{Wasserstein distance}
\label{app:fid_wasserstein}

We use the Wasserstein distance to compare the time-averaged spectra $W(k_y)$ across methods in \cref{tab:downstream_spectral}. It measures the minimum cost of transporting probability mass between two distributions, with cost proportional to the distance the mass is moved, and stays well-defined when the two supports do not overlap, which is useful when generated and reference spectra peak at different $k_y$. We normalise each spectrum so that its total sum is one, ensuring it represents a probability distribution over $k_y$ before computing the distance. Formally,
\[
W_p(P, Q) \;=\; \left( \inf_{\gamma \in \Gamma(P, Q)} \int \|x - y\|^p \, d\gamma(x, y) \right)^{1/p},
\]
where $\Gamma(P, Q)$ is the set of couplings of $P$ and $Q$. We use $p = 1$ on the 1D index $k_y$. \ourfid{} (\cref{sec:fid_pres}, \cref{eq:gyroswin_fid}) is itself a Wasserstein-2 distance between Gaussian fits of the GyroSwin activations.

\subsection{Spectral RMSE and Wasserstein distance}
\label{app:fid_spectral}

\Cref{tab:downstream_spectral} complements the Pearson-correlation columns of \cref{tab:downstream} with two additional spectral fidelity metrics: pointwise RMSE on the unnormalised amplitudes of $\langle W(k_y)\rangle$ and $\langle Q(k_y)\rangle$, and Wasserstein distance on each (computed after normalising every spectrum to a probability distribution over $k_y$, see \cref{app:fid_wasserstein}). RMSE penalises mismatched spectral magnitudes, while the Wasserstein distance is invariant to overall scale and instead penalises misplaced peaks.

\begin{table}[h]
\centering
\caption{Spectral fidelity on in-distribution (\textbf{ID}) and out-of-distribution (\textbf{OOD}) operating conditions. We report RMSE and Wasserstein distance on the time-averaged binormal potential spectrum $W(k_y)$ and flux spectrum $Q(k_y)$. Quasilinear and GPR baselines do not produce $W(k_y)$ pointwise estimates and only provide flux observables. Best \textbf{bold}, second \underline{underlined}.}
\vspace{1em}
\label{tab:downstream_spectral}
\setlength{\tabcolsep}{5pt}
\renewcommand{\arraystretch}{1.1}
\small
\resizebox{\textwidth}{!}{
\begin{tabular}{l cc cc cc cc}
\toprule
\multirow{2}{*}{\textbf{Method}}
 & \multicolumn{2}{c}{$W(k_y)_\text{RMSE}\!\downarrow$}
 & \multicolumn{2}{c}{$Q(k_y)_\text{RMSE}\!\downarrow$}
 & \multicolumn{2}{c}{$W(k_y)_\text{WD}\!\downarrow$}
 & \multicolumn{2}{c}{$Q(k_y)_\text{WD}\!\downarrow$} \\
\cmidrule(lr){2-3} \cmidrule(lr){4-5} \cmidrule(lr){6-7} \cmidrule(lr){8-9}
 & \textbf{ID} & \textbf{OOD} & \textbf{ID} & \textbf{OOD} & \textbf{ID} & \textbf{OOD} & \textbf{ID} & \textbf{OOD} \\
\midrule
Quasilinear                   & $7.37_{\pm 5.66}\!\times\!10^{2}$                                         & $4.22_{\pm 2.61}\!\times\!10^{2}$                 & $\underline{7.38}_{\pm 2.997}$ & $\underline{7.60}_{\pm 3.58}$ & $0.0171_{\pm 0.0152}$            & $0.0135_{\pm 0.0117}$             & $0.0155_{\pm0.007}$                              & $0.0147_{\pm 0.005}$                            \\
GPR                           & ---                                           & ---                                           & ---                            & ---                           & ---                              & ---                               & ---                               & ---                               \\
\midrule
$\text{GyroSwin (warm)}^\ast$ & $6.78_{\pm 6.45}\!\times\!10^{5}$             & $1.60_{\pm 1.48}\!\times\!10^{6}$             & $9.28_{\pm 6.33}$              & $10.53_{\pm 6.34}$            & $\underline{0.010}_{\pm 0.0076}$ & $0.0140_{\pm 0.0061}$             & $\mathbf{0.0065}_{\pm 0.0013}$    & $\mathbf{0.0120}_{\pm 0.0075}$    \\
GyroSwin (cold)               & $2.68_{\pm 2.47}\!\times\!10^{6}$             & $3.51_{\pm 3.44}\!\times\!10^{7}$             & $13.60_{\pm 5.82}$             & $13.87_{\pm 7.71}$            & $0.014_{\pm 0.0103}$             & $0.0162_{\pm 0.0043}$             & $0.0222_{\pm 0.0100}$             & $\underline{0.0129}_{\pm 0.0031}$ \\
\midrule
VAE                           & $4.28_{\pm 3.90}\!\times\!10^{5}$             & $1.15_{\pm 0.99}\!\times\!10^{6}$             & $13.84_{\pm 5.87}$             & $14.07_{\pm 7.40}$            & $0.0139_{\pm 0.0101}$            & $0.0159_{\pm 0.0047}$             & $0.0274_{\pm 0.0075}$             & $0.0266_{\pm 0.0082}$             \\
VQ-VAE (rand)                 & $1.01_{\pm 0.90}\!\times\!10^{5}$             & $\underline{2.85}_{\pm 2.63}\!\times\!10^{5}$ & $12.94_{\pm 6.07}$             & $12.70_{\pm 7.22}$            & $0.0126_{\pm 0.0092}$            & $0.0149_{\pm 0.0047}$             & $0.0257_{\pm 0.0079}$             & $0.0251_{\pm 0.0071}$             \\
VQ-VAE (AR)                   & $\underline{7.87}_{\pm 6.71}\!\times\!10^{4}$ & $3.18_{\pm 2.98}\!\times\!10^{5}$             & $9.94_{\pm 5.07}$              & $8.31_{\pm 5.08}$             & $0.0105_{\pm 0.0080}$            & $\underline{0.0131}_{\pm 0.0057}$ & $0.0136_{\pm 0.0052}$             & $0.0133_{\pm 0.0056}$             \\
\textbf{\ourmethod} (probe)   & $\mathbf{1.71}_{\pm 1.14}\!\times\!10^{3}$    & $\mathbf{6.89}_{\pm 4.67}\!\times\!10^{2}$    & $12.16_{\pm 6.16}$             & $12.34_{\pm 7.92}$            & $\mathbf{0.0093}_{\pm 0.0035}$   & $\mathbf{0.0083}_{\pm 0.0032}$    & $0.0175_{\pm 0.0065}$             & $0.0180_{\pm 0.0080}$             \\
\textbf{\ourmethod} (decode)  & $4.39_{\pm 3.65}\!\times\!10^{5}$             & $3.57_{\pm 3.36}\!\times\!10^{6}$             & $\mathbf{3.37}_{\pm 1.54}$     & $\mathbf{6.75}_{\pm 3.50}$    & $0.0123_{\pm 0.0104}$            & $0.0149_{\pm 0.0050}$             & $\underline{0.0077}_{\pm 0.0021}$ & $0.0137_{\pm 0.0041}$             \\
\bottomrule
\end{tabular}
}
\end{table}

\subsection{Latent depths}

The frozen GyroSwin encoder $g_{\xi}$ of \cref{sec:fid_pres} is a hierarchical Swin5D U-Net composed of three down-up sampling stages with skip connections, a Transformer bottleneck, and an auxiliary decoder that reconstructs the electrostatic potential $\phi$. We extract activations at four depths.
\begin{itemize}[leftmargin=1.5em,itemsep=2pt,topsep=2pt]
  \item \textbf{skip $L{=}2$}: the bottleneck-adjacent middle block.
  \item \textbf{$\phi$ decoder}: the bottleneck-adjacent activation taken from the auxiliary $\phi$-reconstruction branch rather than the $f$-reconstruction branch.
  \item \textbf{flux head}: the activation immediately preceding GyroSwin's scalar flux readout.
\end{itemize}
Skip $L{=}2$, the flux head, and the $\phi$ decoder constitute the headline of \cref{tab:warmstart_fid}\subref{tab:fid_panel}.

\subsection{Sample-count regime}

\ourfid{} of \cref{eq:gyroswin_fid} is the squared 2-Wasserstein distance between Gaussian fits to two empirical activation clouds. The empirical estimator is biased: with $K$ generated and $N_{\mathrm{ref}}$ reference samples and latent dimension $D$, the bias of $\widehat{\Sigma}$ scales like $D/\min(K, N_{\mathrm{ref}})$ and the bias of the trace term in \cref{eq:gyroswin_fid} therefore decays as $\mathcal{O}(1/\min(K, N_{\mathrm{ref}}))$ with a generator-dependent prefactor \citep{Binkowski2018DemystifyingMMD,Chong2020UnbiasedFID}. Standard image benchmarks resolve this by sampling $N_{\mathrm{ref}} \geq 50{,}000$ \citep{Heusel2017FID}; in our setting $K = 64$ and $N_{\mathrm{ref}} \sim 80$, so the absolute values of \ourfid{} are biased and serve only as relative rankings at fixed depth, not as estimates of the underlying $W_2^{2}$.

\newpage

\end{document}